\documentclass[twocolumn]{aastex63}

\usepackage{subfigure}

\shorttitle{CHILES Variables}
\shortauthors{Sarbadhicary et al.}

\graphicspath{{./Plots/}}

\begin{document}

\title{CHILES VERDES: Radio variability at an unprecedented depth and cadence in the COSMOS field}

\correspondingauthor{Sumit K.\ Sarbadhicary}
\email{sarbadhi@msu.edu}

\newcommand{\UNC}{\affiliation{Department of Physics and Astronomy, University of North Carolina at Chapel Hill, Chapel Hill, NC 27599, USA}}
\newcommand{\MSU}{\affiliation{Center for Data Intensive and Time Domain Astronomy, Department of Physics and Astronomy, Michigan State University, East Lansing, MI 48824, USA}}
\newcommand{\LESIA}{\affiliation{LESIA, Observatoire de Paris, CNRS, PSL, SU/UPD, Meudon, France}}
\newcommand{\USydney}{\affiliation{Sydney Institute for Astronomy, School of Physics, University of Sydney, Sydney, New South Wales 2006, Australia}}
\newcommand{\UWMad}{\affiliation{Department of Astronomy, University of Wisconsin-Madison, Madison, WI 53706, USA}}
\newcommand{\NRAO}{\affiliation{National Radio Astronomy Observatory, P.O. Box O, Socorro, NM 87801, USA}}
\newcommand{\UOxford}{\affiliation{Astrophysics, Department of Physics, University of Oxford, Keble Road, Oxford, OX1 3RH, UK}}
\newcommand{\UCapeTown}{\affiliation{Department of Astronomy, University of Cape Town, Private Bag X3, Rondenbosch, 7701, South Africa}}
\author[0000-0002-4781-7291]{Sumit K.\ Sarbadhicary}
\MSU

\author[0000-0002-4039-6703]{Evangelia Tremou}
\LESIA

\author[0000-0001-8026-5903]{Adam J. Stewart}
\USydney

\author[0000-0002-8400-3705]{Laura Chomiuk}
\MSU 

\author[0000-0001-5826-6803]{Charee Peters}
\UWMad

\author[0000-0002-3733-2565]{Chris Hales}
\NRAO

\author[0000-0002-1468-9668]{Jay Strader}
\MSU

\author[0000-0003-3168-5922]{Emmanuel Momjian}
\NRAO

\author{Rob Fender}
\UOxford
\UCapeTown

\author{Eric M. Wilcots}
\UWMad

\begin{abstract}
Although it is well-established that some extragalactic radio sources are time-variable, the properties of this radio variability, and its connection with host galaxy properties, remain to be explored---particularly for faint sources. Here we present an analysis of radio variable sources from the CHILES Variable and Explosive Radio Dynamic Evolution Survey (CHILES VERDES)---a partner project of the 1.4 GHz COSMOS \ion{H}{1} Large Extragalactic Survey (CHILES). CHILES VERDES provides an unprecedented combination of survey depth, duration, and cadence, with 960 hrs of 1--2 GHz continuum VLA data obtained over 209 epochs between 2013 and 2019 in a 0.44 deg$^2$ section of the well-studied extragalactic deep field, COSMOS. We identified 18 moderate-variability sources (showing $10-30\%$ flux density variation) and 40 lower variability sources (2-10$\%$ flux density variation). They are mainly active galactic nuclei (AGN) with radio luminosities in the range of $10^{22}-10^{27}$  W Hz$^{-1}$ based on cross-matching with COSMOS multi-wavelength catalogs.
The moderate-variability sources span redshifts $z=0.22-1.56$, have mostly flat radio spectra ($\alpha>-0.5$), and vary on timescales ranging from days to years.  Lower-variability sources have similar properties,
but have generally higher radio luminosities than the moderate-variability sources,
extend to $z = 2.8$, and have steeper radio spectra ($\alpha<-0.5$). No star-forming galaxy showed statistically significant variability in our analysis. The observed variability likely originates from scintillation on short ($\sim$week) timescales, and Doppler-boosted intrinsic AGN variability on long (month--year) timescales.

\end{abstract}


\section{Introduction} \label{sec:intro}
The time-domain radio sky hosts a diversity of astrophysical phenomena, 
including Galactic events such as novae, flare stars, X-ray binaries, cataclysmic variables and magnetar flares, and extragalactic phenomena like supernovae, gamma-ray bursts, tidal-disruption events, neutron star mergers and active galactic nucleus (AGN) flares \citep{Metzger2015, Fender2017}. The most common phenomenon in time-domain radio surveys is AGN variability \citep{Thyagarajan2011, Mooley2016}, which is hypothesized to be caused by shocks within a relativistic jet powered by an accreting supermassive black hole \citep{Marscher1985, Hughes1989}. 

Radio observations complement other wavelengths in the characterization of these events for several reasons. Radio directly probes shocks and relativistic jets via synchrotron emission \citep{Chevalier2017, Panessa2019}, is sensitive to dust-obscured phenomena missed in optical wavebands \citep{GalYam2006, Brunthaler2009}, and can provide fast, accurate localization of events compared to wavelengths like X-ray and gamma-rays. Monitoring the variable radio sky is therefore one of the main science drivers of future radio facilities such as the Square Kilometer Array \citep[SKA,][]{Fender2015} and the Next Generation Very Large Array \citep[ngVLA,][]{Murphy2018}.

The last decade has seen a rise in the number of blind surveys attempting to systematically discover and characterize the rates of radio-variability in the sky\footnote{For an exhaustive list, see \url{http://www.tauceti.caltech.edu/kunal/radio-transient-surveys/}}. These efforts were motivated by the emergence of powerful radio facilities with increasingly sensitive wide-field and wide-band observing capabilities such as the Karl G.\ Jansky Very Large Array (VLA; \citealt{Perley2011}), 
MeerKAT \citep{Booth2012}, Australian SKA Pathfinder Telescope \citep[ASKAP,][]{Murphy2013}, and low-frequency facilities like the Low Frequency Array \citep[LOFAR,][]{vanHarlem2013}, and Murchison Widefield Array \citep[MWA,][]{Bell2019}. Studies have converged on a common result: the time-variable radio sky is relatively quiet. Only a handful of transients have been discovered in blind surveys \citep[e.g.][]{Gregory1986, Levinson2002, Bower2007, Bannister2011, Jaeger2012, Mooley2016, Radcliffe2019, Driessen2020}. Most radio transient surveys have yielded non-detections, and found that only about 1$\%$ of the radio source population is variable \citep{Carilli2003, Frail2003, deVries2004, Lazio2010, Bower2010, Bower2011a, Croft2010, Croft2011, Croft2013, Bell2011, Ofek2011, Thyagarajan2011}. 

Time-domain radio surveys face several challenges. First-generation radio surveys mainly probed radio sources brighter than 1 mJy, and relied often on archival all-sky datasets such as NRAO VLA Sky Survey \citep[NVSS,][]{Condon1998} and the Faint Images of the Radio Sky at Twenty Centimeters Survey \citep[FIRST,][]{Becker1995}. The mismatched sensitivities, spatial resolutions and absolute flux density calibrations encountered when comparing these surveys made it challenging to produce a complete sample of variables and transients. Imaging artifacts as a result of improper data cleaning, calibration and sidelobe contamination were also identified as a potential source of false-positive detections \citep{Frail2012}. More recent studies have utilized the newer generation of sensitive radio facilities, deep-field extragalactic observations, and/or improvements in mosaicking and imaging techniques \citep{Mooley2013, Mooley2016, Hancock2016, Bhandari2018, Radcliffe2019}, but are still forced to work with just a few epochs as a compromise between epoch sensitivity, total time allocation, and survey area. The small number of epochs makes it difficult to characterize the temporal behavior of transients and variables occurring on a wide variety of timescales in the radio sky \citep{Metzger2015}.

Here, we fill in a portion of this under-explored parameter space with
the CHILES Variable and  Explosive Radio Dynamic Evolution Survey (CHILES VERDES), designed to look for variable phenomena within a single VLA pointing observed at L-band (1--2 GHz),
using nearly 1000 hours of observations obtained between 2013 and 2019. The data was taken as part of the COSMOS \ion{H}{1} Legacy Survey (CHILES), a survey of neutral hydrogen traced by the 21 cm emission line in the COSMOS field out to an unprecedented depth of $z\approx0.5$ \citep{Fernandez2013, Dodson2016, Fernandez2016, Hess2019, BlueBird2019, Luber2019}. The VLA Wideband Interferometric Digital ARchitecture (WIDAR) correlator also provides simultaneous wide-band continuum observations, which can be searched for synchrotron emission from energetic phenomena. 

The CHILES VERDES survey is unique amongst GHz-radio variable surveys with its unprecedented combination of survey depth, cadence and duration---sources are sampled with 172 epochs of observations every few days for 5.5 years (except for gaps between B-configuration semesters), with each epoch reaching RMS sensitivities of $\sim 10\,\mu$Jy beam$^{-1}$ per epoch. While the 22.5$^{\prime}$ radius field of view limits the recovered sample size, we are able to probe down to faint flux densities ($<$100 $\mu$Jy),
and sample the time-domain characteristics of our sources on timescales spanning days to years. The survey also avoids the challenges associated with mosaics and multiple pointings \citep[e.g.][]{Mooley2013, Mooley2016}, and thus complements shallower, wide-field variability surveys. 
Finally, the COSMOS field is well supplemented with multi-wavelength data, providing us redshifts, multi-wavelength luminosities, and detailed characterization of the host galaxies for our sample \citep{Scoville2007}.

In this paper, we present a survey of variables carried out with the CHILES VERDES observations. A separate paper will discuss the discovery and statistics of transient phenomena in CHILES VERDES (Stewart et al., in prep). AGN variables are the dominant category of events discovered in time-domain radio surveys \citep{Thyagarajan2011}, vary on timescales similar to transients \citep{Metzger2015, Mooley2016}, and can be misinterpreted as transients \citep[e.g.,][]{Williams2016}. With more sensitive blind surveys, variables will likely dominate the background or confusion-level against which transients are discovered \citep{Carilli2003, Rowlinson2019}. It is therefore important to understand the variability characteristics of the radio source population in blind surveys. Particularly important, and relatively less explored, is variability in the sub-mJy regime, where the source population becomes more dominated by star-forming galaxies \citep{Smolcic2017b} and the expected rate of all categories of transients is higher \citep{Frail2012, Metzger2015, Mooley2013}. The few surveys is this regime \citep[e.g.,][]{Carilli2003, Radcliffe2019} show that the variable fraction in the sub-mJy regime can be larger than in the mJy regime. Extagalactic variability is also an important constraint on scintillation models and turbulence in the ionized interstellar medium (ISM), particularly off the Galactic plane where pulsars are rare \citep{Rickett1990}. 

This paper is organized as follows: \S \ref{sec:chiles} describes the CHILES VERDES observations, data reduction and imaging, \S \ref{sec:trap} describes the search method and analysis of variables using an automated pipeline, and \S \ref{sec:chilesvarcandidates} describes the statistical properties, light curves and structure functions of the radio variables. \S \ref{sec:multiwav} discusses the host galaxy properties, AGN properties, redshift distribution, and spectral indices of our sources from multi-wavelength catalogs. In \S \ref{sec:discuss}, we discuss the information learned about radio variability from our analyses, and we conclude in \S \ref{sec:conclusion}.

\section{The Survey} \label{sec:chiles}
\subsection{The CHILES survey}

The COSMOS \ion{H}{1} Legacy Extragalactic Survey, or CHILES, is an ambitious project using the VLA to obtain a deep field in the \ion{H}{1} 21 cm spectral line \citep{Fernandez2013, Fernandez2016}. The enormous correlator power of the upgraded (Jansky) VLA allows observers to obtain large bandwidths at high spectral resolution \citep{Perley2011}, enabling a survey for neutral hydrogen from $z = 0.0-0.5$ in a single observation. 

The CHILES pointing was chosen to be in the COSMOS field \citep{Scoville2007}, where it can enjoy significant multi-wavelength supporting data (e.g., \citealt{Davies2015, Civano2016, Andrews2017}). It is centered at J2000 position, RA = $10h01m24.00s$, Dec = $+02^{\circ}21^{\prime}00.0^{\prime\prime}$. The single L-band pointing yields a field of view 0.5$^{\circ}$ in diameter (half power beam width) at 1.5 GHz (we note that our search diameter for variables is slightly larger, see \S \ref{sec:chilesvarcandidates}).
Data were obtained in the VLA's B-configuration, which yields a resolution of $4.3^{\prime\prime}$ at 1.5 GHz.

The CHILES data was obtained in both spectral line and continuum data,
the latter observed in full polarization mode and comprised of four spectral windows, each with 128 MHz bandwidth sampled by 64 channels. These spectral windows span L-band and cover frequencies relatively free of RFI, centered at 1032, 1416, 1672, and 1800 MHz. CHILES data were saved to disk with 8s correlator integration time. Data were calibrated using 3C286 as a bandpass and absolute flux density calibrator and J0943$-$0819 as a complex gain calibrator.  

\subsection{The CHILES VERDES Project} \label{sec:chilesverdes}

The CHILES VERDES project is the study of transient and variable sources within the CHILES deep field, using the continuum dataset produced by CHILES. 
Data were observed over the time frames listed in Table \ref{log}. Each VLA B-configuration semester lasts approximately four months, and is followed by a 12 month break before the VLA re-enters B-configuration.
During a given B-configuration semester, CHILES observations were obtained, on average, every few days. CHILES observations were obtained in scheduling blocks of durations ranging from 1--8 hr, with a median duration of 4.5 hr. We call each of these scheduling blocks an ``epoch" throughout this paper.

\begin{deluxetable*}{lcccc}
\tablecaption{ \label{log}
Log of CHILES VERDES Observations}
\tablehead{Semester & Date Range & Number of Epochs & Total Time (hr) & Total Time on Source (hr)}
\startdata
1 & 25 Oct 2013--21 Jan 2014 & 45 &  169.55 & 128.56  \\
2 & 25 Feb 2015--04 May 2015 & 39 & 207.11 & 155.78\\
3 & 16 May 2016--27 Sep 2016 & 49 & 177.52 & 133.36\\
4 & 11 Nov 2017--29 Jan 2018 & 48 & 227.51 & 170.77\\
5 & 01 Mar 2019--11 Apr 2019 & 28 & 176.78 & 134.05\\
\enddata
\end{deluxetable*} 
The continuum data were reduced using a custom-developed scripted pipeline for the CHILES Continuum Polarization (or CHILES Con Pol) project, operated in CASA release version 4.7.2. Unlike CHILES VERDES which uses total-intensity (Stokes I) continuum maps, CHILES Con Pol is a partner project dedicated to making full-polarization maps of the CHILES data in order to probe the 
cosmological evolution of polarized radio sources and intergalactic magnetic fields \citep{Hales2014b}\footnote{\url{http://www.chilesconpol.com/}}. The pipeline was run in a semi-automated manner by a single person (C.\ Hales), and each calibration step was manually inspected for quality assurance. High quality flagging was performed using \texttt{pieflag} \citep{Hales2014a}, supplemented by CASA functionality provided by the custom tasks \texttt{antintflag3} \citep{Hales2016a}, \texttt{plot3d} \citep{Hales2016b}, and \texttt{interpgain} \citep{Hales2016c}. An additional assessment of calibration quality for each observation was carried out in \cite{Hales2019} by applying the calibration solutions to 3C286, measuring the spectrum of fractional linear polarization, and comparing this with the known spectrum from D configuration observations by \citet{Perley2013a}. 
The pipeline was run twice for each observation, the first time to obtain calibrated data
and a quick-look image within a few hours of coming off the telescope (for transient searches and potential optical followup), and the second at least 2 weeks after the observation so as to incorporate GPS-derived ionospheric total electron content (TEC) data obtained from the International Global Navigation Satellite System Service (IGS), which has a latency period of up to 2 weeks. Our variable search described in \S \ref{sec:trap} was carried out on the TEC-corrected total-intensity data.

We create a time-averaged Stokes I image for each epoch using \verb|tclean| in CASA \citep{McMullin2007}. We use multi-frequency synthesis with \texttt{nterms = 2} and a reference frequency of 1.45 GHz.
 Each image is substantially larger than the primary beam half-power point, and calculated with w-projection and \texttt{wprojplanes = 128}, but is truncated at a radius corresponding to 20\% of the sensitivity at image center. Images are created with Briggs weighting of \texttt{robust=0.7}, and pixels are 1$^{\prime\prime}$ across. Image deconvolution was carried out with the multi-scale multi-frequency synthesis algorithm on scales of [0,5,15] pixels \citep{Rau2011}, and a clean threshold was set at $6\times$ the image theoretical noise.
 
\begin{figure*}
    \centering
    \includegraphics[width=\textwidth]{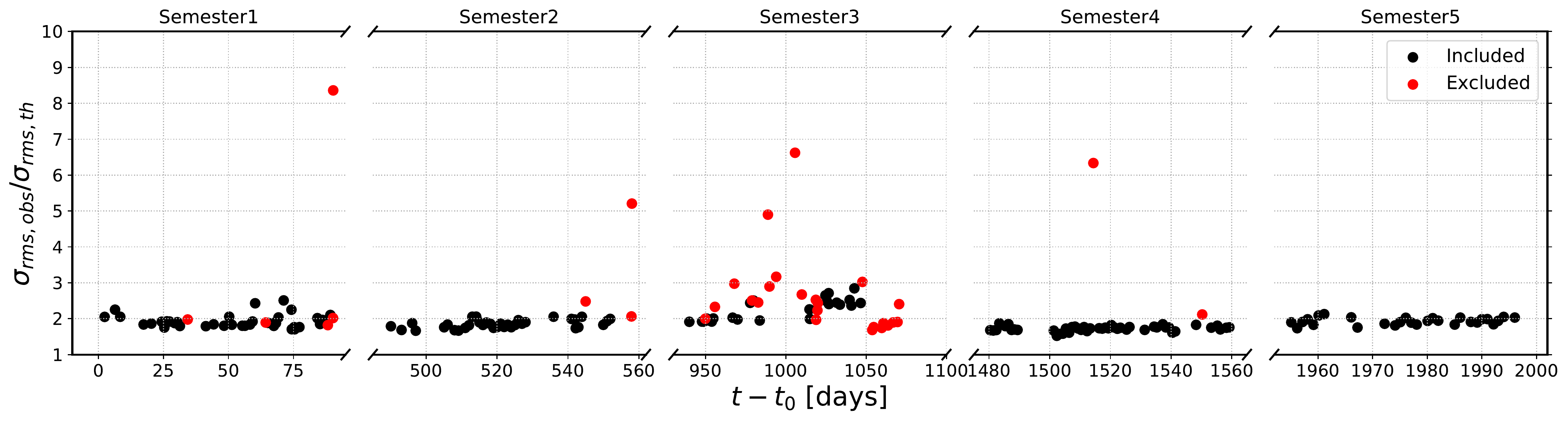}
    \caption{Ratio of the image RMS noise ($\sigma_{rms,obs}$) to the theoretical RMS noise ($\sigma_{rms,th}$) as a function of time for all 209 CHILES VERDES epochs, divided by B-configuration semester. The 172 included epochs and the 38 excluded epochs
    are shown as black and red markers respectively. The time-axis is the CHILES epoch in units of days starting from our first observation, $t_0$ = 2456588.5 JD (2013 Oct 23).}.
    \label{fig:rms}
\end{figure*}

A total of 209 CHILES continuum images (totaling 960 hours of observations) were obtained between 2013 and 2019 from all the B-configuration scheduling blocks, of which we decided to use 172 for our CHILES VERDES search. The images have a typical spatial resolution of 4.5$^{\prime\prime}$, with a range of $3.8^{\prime\prime}-5.7^{\prime\prime}$ from image to image. The typical non-primary beam-corrected image RMS noise per epoch is about 10$\mu$Jy beam$^{-1}$, varying from image to image within the range 7--27 $\mu$Jy beam$^{-1}$ depending on the on-source integration time per epoch. The RMS noise was calculated inside a relatively uncrowded region of the CHILES field using the CASA task \texttt{imstat}. Of the rejected images, nine were obtained during the VLA B$\rightarrow$A move time at the end of the 2016 semester, resulting in a systematic decrease in intensity for many CHILES VERDES sources due to the changing $uv$-coverage. The rest of the rejected images had various issues, from high RMS noise to residual artifacts in the final images, that made them unsuitable for probing variability. The rejected images were affected either by corrupted data, high RFI levels due to daytime observations (particularly in 2016), or sub-optimal calibration solutions obtained by the pipeline on short ($\sim$1 hr) epochs. A more detailed investigation will be carried out in the future to recover these epochs. For our purposes, the current set of 172 images have stable image quality, with image RMS being roughly a factor of 2 of the theoretical RMS noise values in Figure \ref{fig:rms}. The theoretical RMS noise was calculated using the VLA sensitivity equation\footnote{\url{https://science.nrao.edu/facilities/vla/docs/manuals/oss/performance/sensitivity}}, given the number of antennas used, total time on source, and an assumed bandwidth of 0.5 GHz.

\section{Finding Variables with the LOFAR Transients Pipeline}  \label{sec:trap}
We identify potential sources and extract their light curves from our radio images using the LOFAR Transients Pipeline (TraP; \citealt{Swinbank2015}). TraP can systematically extract radio sources across multiple images using a set of uniform and quantifiable selection criteria, and has been used by a variety of contemporary radio variability surveys \citep[e.g.,][]{Hobbs2016, Stewart2016, Rowlinson2019, Driessen2020}.

The details of the TraP process is described in \cite{Swinbank2015}, but for the convenience of the reader, we provide a brief description of the process. We ran TraP on our non-primary-beam corrected images (excluding the rejected images described in \S \ref{sec:chilesverdes}), which have more uniform noise properties than primary-beam corrected images and provide the opportunity to discover interesting variable sources farther away from the phase center. Once the final source catalog is obtained, primary beam correction is applied to the flux densities of sources (\S \ref{sec:pbcorrect}). The first image in the time series searched by TraP is a deep reference image obtained by co-adding the 2013 semester data. This was done so TraP can reliably identify sources with accurate positions, and to prevent spurious transient or variable detections from sources near the detection thresholds. TraP identified sources in each image by selecting pixels with flux densities above a 5$\sigma$ threshold compared to the local RMS,
and then ``growing" the source region to adjacent pixels above a 3$\sigma$ threshold value.
For each of these sources, TraP records the position, flux densities, and their respective uncertainties per image. TraP calculated the flux densities of the sources by fitting an elliptical Gaussian at the recovered source location, keeping the width of the fitting beam fixed to the shape of the synthesized beam. It then matches each detected source per image with its counterpart in adjacent images using the dimensionless de-Ruiter radius ($r_{ij}$), which is the angular separation between source $i$ and its counterpart $j$, normalized by their position uncertainties \citep{deRuiter1977}. A source is said to have a counterpart in the subsequent image if their angular separation is
less than the semi-major axis of the restoring beam, and if $r_{ij} \leq r_s$ (where $r_s$ is a threshold set by the \texttt{deruiter\_radius=5.68} so that the probability of missing an associated source $< 10^{-7}$). TraP is able to unambiguously match point-like sources across images, but a small fraction of sources (mostly resolved) get identified as multiple entries in the final source catalog. These entries get automatically filtered out in the variable selection criteria further applied by us in \S \ref{sec:select}.

\subsection{Primary Beam Correction} \label{sec:pbcorrect}
After TraP has searched through all the epochs and recovered a source catalog, we correct the integrated flux density of each source for decreasing sensitivity 
of the primary beam 
with increasing distance from the phase center. We follow the formalism of \cite{Perley2016} by dividing the non-primary-beam-corrected flux densities by a factor $f(x)$ defined as
\begin{equation} \label{eq:pbcor}
    f(x) = 1 + a_1x + a_2x^2 + a_3x^3
\end{equation}
where $x = (R\,\nu_{obs})^2$, $R$ is the distance of the source from the beam center in arcmin, and $\nu_{obs}$ is the observing frequency in GHz. In this paper, we use $a_1 = -1.343 \times 10^{-3}$, $a_2 = 6.579 \times 10^{-7}$ and $a_3 = -1.186 \times 10^{-10}$, which are the values for an average observing frequency of 1.465 GHz.%
\footnote{Primary beam correction formula and coefficients are adopted from \url{http://www.aips.nrao.edu/cgi-bin/ZXHLP2.PL?PBCOR}}
All statistical calculations in the subsequent sections use the primary-beam-corrected flux densities.

\subsection{Identifying Variables} \label{sec:identifyvars}
The sources recovered by TraP will be a mixture of steady (non-variable) and variable sources, and we can differentiate them using statistics calculated from their radio light curves. A radio light curve is composed of a series of flux density measurements $F_i$, each with uncertainty $\sigma_{F,i}$; $N$ is the number of measurements (=172 for most sources).
We use these data to calculate the flux coefficient of variation ($V$) and the weighted reduced $\chi^2$ statistic ($\eta$), calculated as: 
\begin{equation} \label{eq:V}
    V = \frac{s}{\overline{F}} = \frac{1}{\overline{F}} \sqrt{\frac{N}{N-1} \left(\overline{F^2} - \overline{F}^2\right)}
\end{equation}
and
\begin{equation} \label{eq:eta}
    \eta = \frac{1}{N-1} \sum_{i=1}^{N} \frac{\left(F_i - \xi_{F}\right)^2}{\sigma_{F,i}^2}.
\end{equation}
Here, $s$ refers to the standard deviation of the flux density measurements $F_i$, and $\overline{F}$ is the arithmetic mean of $F_i$. 
$\xi_F$ is the weighted mean of the flux density measurements, defined as:
\begin{equation}
    \xi_F = \frac{\sum_{i=1}^{N} F_i/\sigma_{F,i}^2}{\sum_{i=1}^{N} 1/\sigma_{F,i}^2}
\end{equation}
 The coefficient $V$ is equivalent to the fractional variability or modulation index parameter ($m$) used in previous transient surveys \citep[e.g.,][]{Jenet2003, Bell2014, Mooley2016}. The $\eta$ parameter is similar to the reduced $\chi^2$ statistic, and quantifies the significance of the variability.

\begin{figure}
    \centering
    \includegraphics[width=\columnwidth]{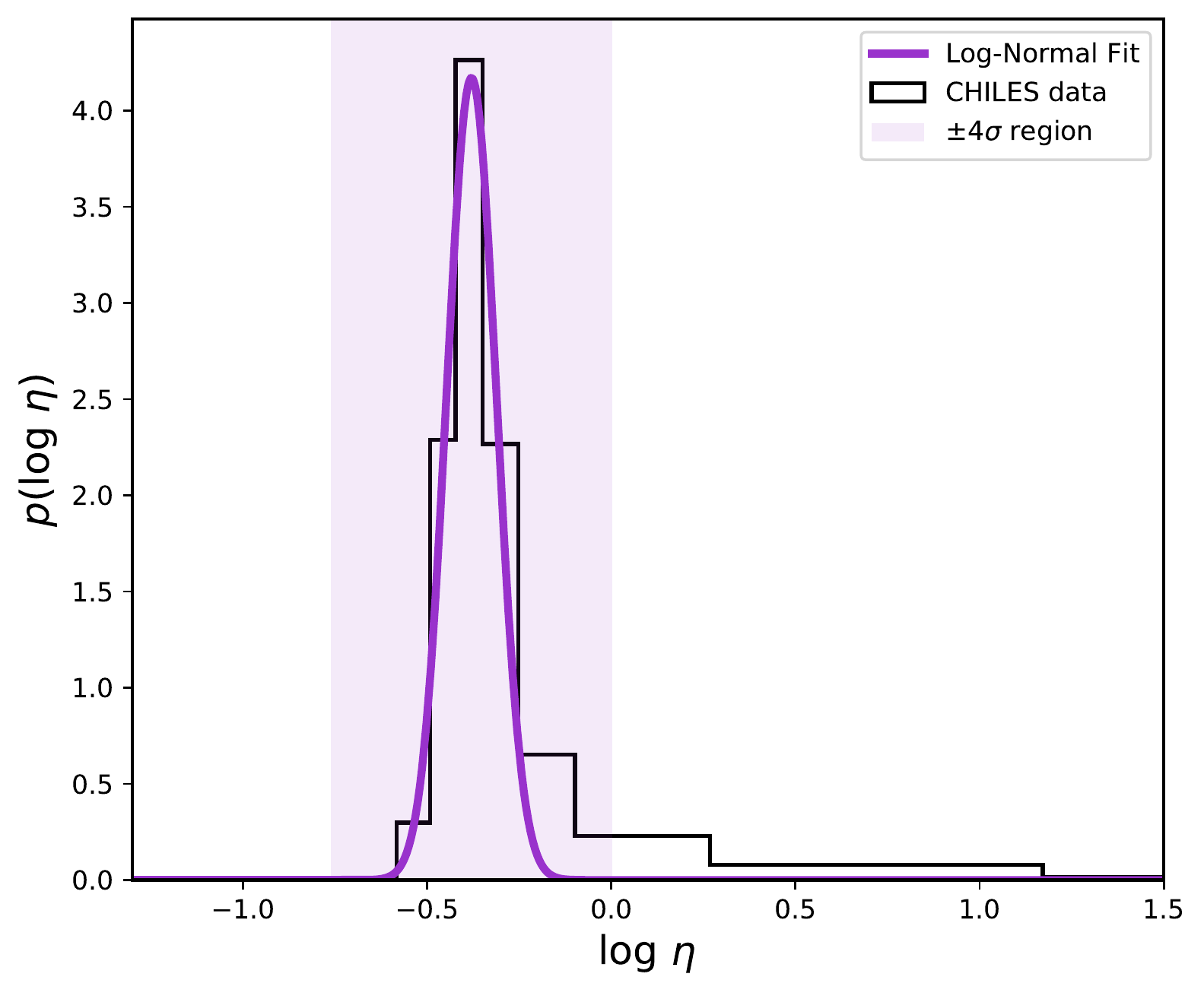}
    \caption{Threshold determination for $\eta$ as described in \S \ref{sec:identifyvars}. The histogram shows the normalized $\mathrm{log}\ \eta$-distribution of our 370 point-like sources. The curve shows the best fit normal function. Sources with $\eta$ within the shaded 4$\sigma$ region are designated steady sources, while those with larger $\eta$ values are designated variable candidates.}
    \label{fig:etathreshold}
\end{figure}

Together, $V$ and $\eta$ has been shown to effectively separate the parameter space of variables from steady sources \citep{Swinbank2015, Rowlinson2019}. A common practice for identifying potential variable sources is to select those with $V$ and $\eta$ above some threshold, defined using the mean and standard deviation of the $V$ and $\eta$ distributions \citep[e.g.,][]{Rowlinson2016, Stewart2016}. We adopt a similar strategy here, 
first determining a threshold on $\eta$ by fitting the $\mathrm{log}\ \eta$ distribution with a normal function, with mean ($\mu_{\mathrm{log}\eta}$) and standard deviation ($\sigma_{\mathrm{log}\eta}$) as free parameters. We follow the strategy of \cite{Rowlinson2019} by making a histogram of the $\mathrm{log}\ \eta$ data with binning determined using the Bayesian Blocks algorithm \citep{Scargle2013} implemented in \texttt{astropy}\footnote{implemented via \texttt{astropy.visualization} in Python 3+, see https://docs.astropy.org/en/stable/visualization/histogram.html}. Bayesian Blocks is part of a family of algorithms that chooses the optimal binning that minimizes the error of the histogram's approximation of the data, with the added advantage of allowing variable bin widths. The histogram is then fitted with a normal function using a non-linear least-squares method\footnote{implemented via \texttt{scipy.optimize.curve\_fit()}} to obtain the best-fit values for $\mu_{\mathrm{log}\eta}$ and $\sigma_{\mathrm{log}\eta}$ (Figure \ref{fig:etathreshold}). 

For this paper, we choose a 4$\sigma$-threshold in $\mathrm{log}\,\eta$. Sources with $\mathrm{log}\,\eta > (\mu_{\mathrm{log} \eta} + 4 \sigma_{\mathrm{log}\eta})$ are considered variable candidates, while sources below this threshold are considered steady sources (Figure \ref{fig:etathreshold}).

\subsection{Structure Functions} \label{sec:sf}
The high signal-to-noise (S/N), well-sampled light curves in CHILES VERDES provide an excellent opportunity to examine the power spectra of radio variability in extragalactic sources, and to do this---possibly for the first time---in a detailed manner with a \emph{blind} radio survey. In an attempt to understand these power spectra, we calculate structure functions (SFs), which have commonly been used to quantify variability power spectra for blazars and quasars \citep[e.g.,][]{Hughes1992}, X-ray binaries \citep[e.g.,][]{Plotkin2019}, and radio scintillation \citep{Simonetti1985}. We apply the SF formalism of \citet[][hereafter P19]{Plotkin2019} to our sources. 

For a light curve with flux density measurements $F(t)$, the first-order SF is defined as 
\begin{equation}
    V(\tau) = \langle \left(F(t+\tau) - F(t)\right)^2 \rangle
\end{equation}
where $\tau$ is defined as the lag-time, or the time difference between two different epochs. For our discretely sampled light curves, we calculate $V(\tau)$ by taking each pair of flux density measurements in the light curve $(F_i, F_j)$ taken at times $(t_i, t_j)$ and calculating $V_{ij} (\tau_{ij}) = (F_j - F_i)^2$, where $\tau_{ij} = t_j - t_i$. Similar to P19, we bin $V_{ij}$ into bins of $\tau_{ij}$, and calculate the mean $V_{ij}$ in that bin. The uncertainty in $V_{ij}$ is given by the standard error $=\sigma/\sqrt{N}$, where $\sigma$ is the standard deviation of the $V_{ij}$ values in the bin and $N$ is the number of points per bin of $\tau_{ij}$. The bin sizes were chosen to be approximately uniformly spaced in log lag-time, although note that because of gaps between semesters, we do not have information on $V(\tau)$ for lag-times between 123 and 363 days. 

As explained in \cite{Hughes1992}, the shape of the SF can reveal information about the underlying physical processes driving variability. At the shortest lag-times, the SF plateaus to roughly twice the variance of flux density uncertainties, and at the longest time-lags to roughly twice the variance of the flux density measurements. The shape of the SF in between these extremes depend on the noise regime \citep[e.g., white noise, flicker noise, shot noise;][]{Hughes1992}. The shortest time-lag at which the SF breaks from the white noise regime (where the SF flattens) corresponds to the shortest timescale of uncorrelated behavior (in the case of white noise) or the distribution of impulse response times (in the case of flicker noise). It is generally interpreted as the characteristic timescale of the underlying process driving variability.

We acknowledge however that shapes of SFs are affected by the finite length of light curves, large gaps in observations and the random nature of variability, and therefore may not accurately represent the true variability power spectrum of the source \citep{Emmanoulopoulos2010}. Furthermore in many of the variables in Figures \ref{fig:rlcsfvars1}, \ref{fig:rlcsfvars2} and \ref{fig:rlcsflowvars}, the SFs have complex shapes with no well-defined plateaus or breaks, and cannot be easily fit with simple functional forms to derive characteristic timescales. We therefore defer a more detailed analysis of the variability spectrum for future papers, and simply present the SFs alongside the light curves as a rough guide for the potential timescales present in the light curves. 

\section{CHILES VERDES Variable Candidates} \label{sec:chilesvarcandidates}
\subsection{Selecting candidates} \label{sec:select}
\begin{figure}
    \centering
    \includegraphics[width=\columnwidth]{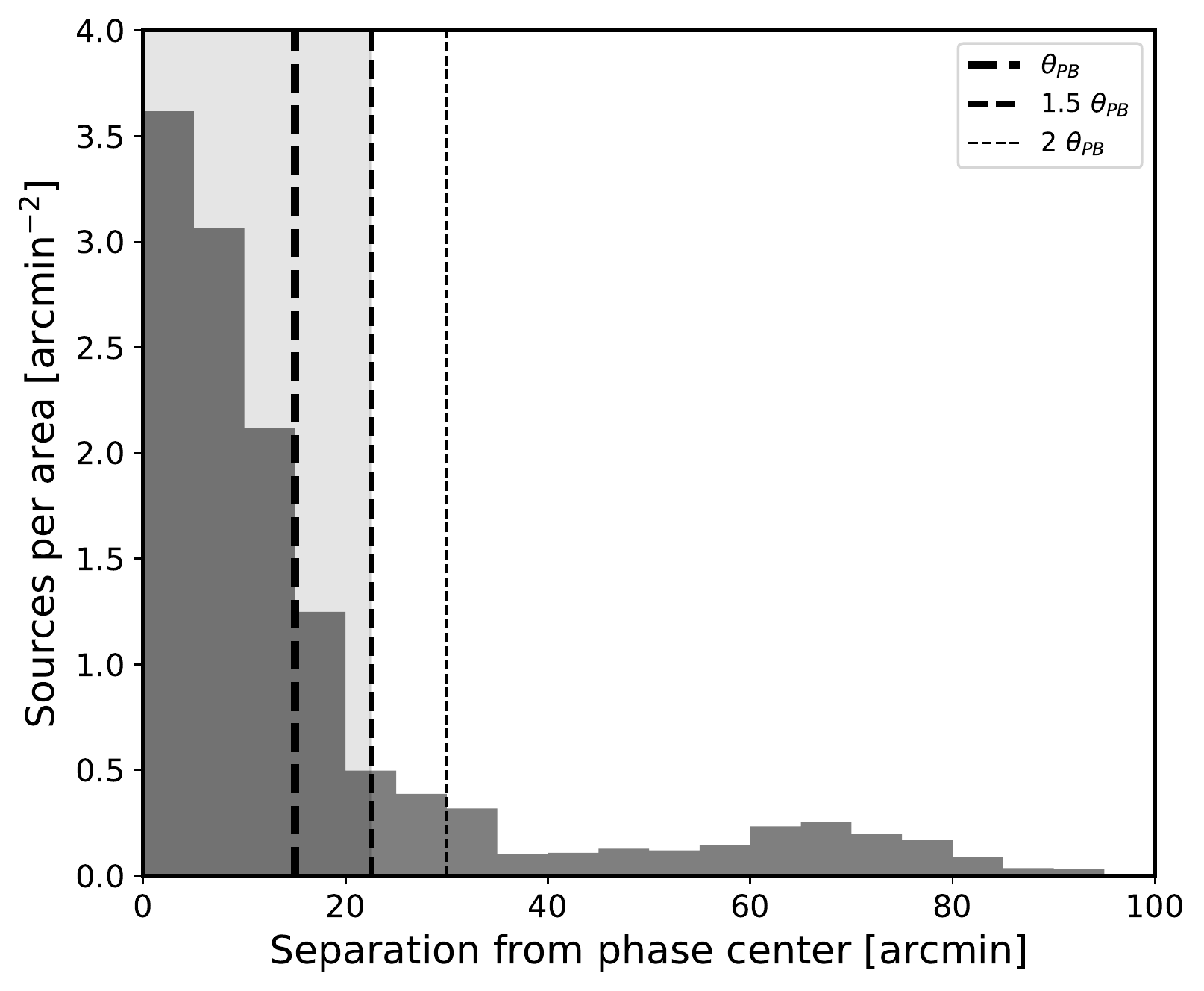}
    \caption{Surface density of sources found by TraP within the non-pbcor images, versus their separation from the phase center. The vertical lines show different distance cuts from the L-band primary beam phase center in units of the half-width at half-maximum, $\theta_{PB} = 15^{\prime}$ at 1.5 GHz. Since the surface density begins to flatten roughly at 1.5~HWHM$_{PB}$, we include all sources within this radius for our variability analysis.}
    \label{fig:sep}
\end{figure}
\begin{figure}
    \centering
    \includegraphics[width=1.12\columnwidth]{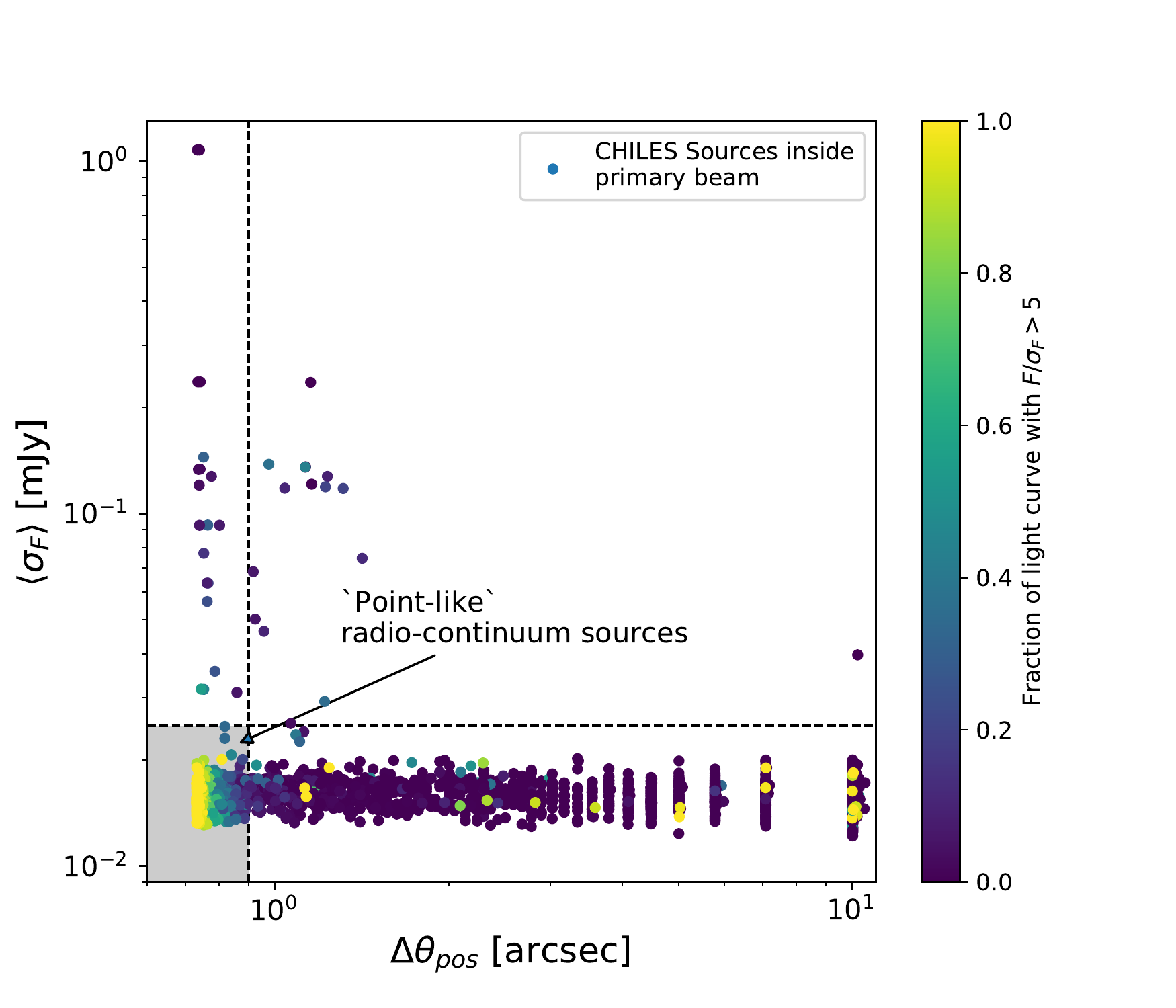}
    \caption{Mean flux density uncertainties ($\langle \sigma_F\rangle$) versus position uncertainties ($\Delta \theta_{pos}$; as defined in \S \ref{sec:select}) for all CHILES sources identified by TraP inside the $1.5 \times\,{\rm HWHM}_{PB}$ area. Sources with $\Delta \theta_{pos} < 0.9^{\prime\prime}$ and $\langle \sigma_I\rangle < 0.025$ mJy (grey region) are considered variable candidates for the remainder of the paper. The colorbar shows the fraction of flux density measurements $>5\times$ the local RMS noise.}
    \label{fig:cuts}
\end{figure}
TraP searched a total of 172 images obtained between 2013 and 2019 that passed quality control, and yielded a catalog of 6821 sources with light curves. The distribution of these sources as a function of distance from the phase center is shown in Figure \ref{fig:sep}. The concentration of sources detected is highest near the phase center, and drops off rapidly as  sensitivity decreases at large radius. For the subsequent analysis of the variable population in CHILES, we restrict ourselves to sources within a radius of 22.5$^{\prime}$, which is $1.5 \times$ the primary beam half-width at half maximum (HWHM) at 1.5 GHz (HWHM$_{PB} = 15^{\prime}$, corresponding to about $17\%$ primary beam sensitivity based on Eq \ref{eq:pbcor}), beyond which the source density undergoes a noticeable flattening. 

We find a total of 2713 sources inside the 1.5~HWHM$_{PB}$ region. Before investigating which of these sources are variable, we carry out two other selection cuts  based on the positional uncertainties ($\Delta \theta_{pos}$) and average flux density uncertainties ($\langle{\sigma_I\rangle}$) returned by TraP, shown in Figure \ref{fig:cuts}. The purpose of these cuts is to select for compact or point-like sources, which are the only sources expected to vary on a $\sim$6 year timescale, and the sources which will have reliable photometry returned from TraP.
\begin{enumerate}
    \item $\Delta \theta_{pos} \leq 0.9^{\prime\prime}$ -- We refer to the uncertainties of the RA and Dec of the sources returned by TraP as $\sigma_{ra}$ and $\sigma_{dec}$ respectively. We define $\Delta \theta_{pos} = \sqrt{\sigma_{ra} \sigma_{dec}}$, the geometric mean of the RA and Dec uncertainties. The average synthesized beam FWHM of our images is about 4.5$^{\prime\prime}$, so for sources observed at S/N = 5, their positional accuracy should be $\sim 4.5^{\prime\prime}/5 \approx 0.9^{\prime\prime}$. Nearly 85$\%$ of the sources in Figure \ref{fig:cuts} have $\Delta \theta_{pos}>0.9^{\prime\prime}$. Visual inspection revealed that the majority of these sources are faint (low S/N), while the rest are associated with bright extended sources with lobes, jets and filaments, whose spatial scales would be too large to produce variability observable on our timescales.  We therefore restrict ourselves to sources with $\Delta \theta_{pos} \leq 0.9^{\prime\prime}$. 
    
    \item $\langle{\sigma_F\rangle} \leq 25$ $\mu$Jy -- The average flux density measurement uncertainties of our sources are narrowly distributed around log$\mathrm{\langle \sigma_F/\mu Jy \rangle} = 1.2 \pm 0.04$ (or $\mathrm{\langle \sigma_F \rangle} \approx 16 \mathrm{\mu}$Jy) as seen in Figure \ref{fig:cuts}, which is a typical RMS noise in the individual epoch images. Visual inspection revealed that sources with high $\langle \sigma_F \rangle$ are mainly associated with bright extended features. We exclude these sources by setting an upper limit of $\langle \sigma_F \rangle < 25 \mu$Jy (roughly 5 times the standard deviation of the log $\langle \sigma_F \rangle$ distribution) for our final catalog.
\end{enumerate} 

We note some caveats here in our selection criteria. It is possible that some fraction of the faint sources that were excluded with $\Delta \theta_{pos}\leq 0.9^{\prime\prime}$ criteria may in fact exhibit variability that can be measured more accurately and reliably in deeper images obtained by co-adding multiple epochs. Such deep images are being prepared as part of our follow-up transients paper (Stewart et al, in prep), where we will comment on them further. Additionally, there may be some compact components of extended sourcs that could exhibit variability, but the surrounding diffuse emission leads them to having a lower signal-to noise over the local background, and therefore skipped by TraP's selection criteria. Making a catalog of such sources might be possible by re-imaging the individual epochs excluding the shorter baselines to filter out extended emission (e.g. from jets, filaments), and re-running TraP on this image-series. We defer additional analysis to a future paper. We note however that such extended sources form a small fraction of the source population in the CHILES field, and is unlikely to change the statistical results we infer for the variable population.

\subsection{$V-\eta$ statistics} \label{sec:varstats}
 \begin{figure*}
    \centering
    \includegraphics[width=0.8\textwidth]{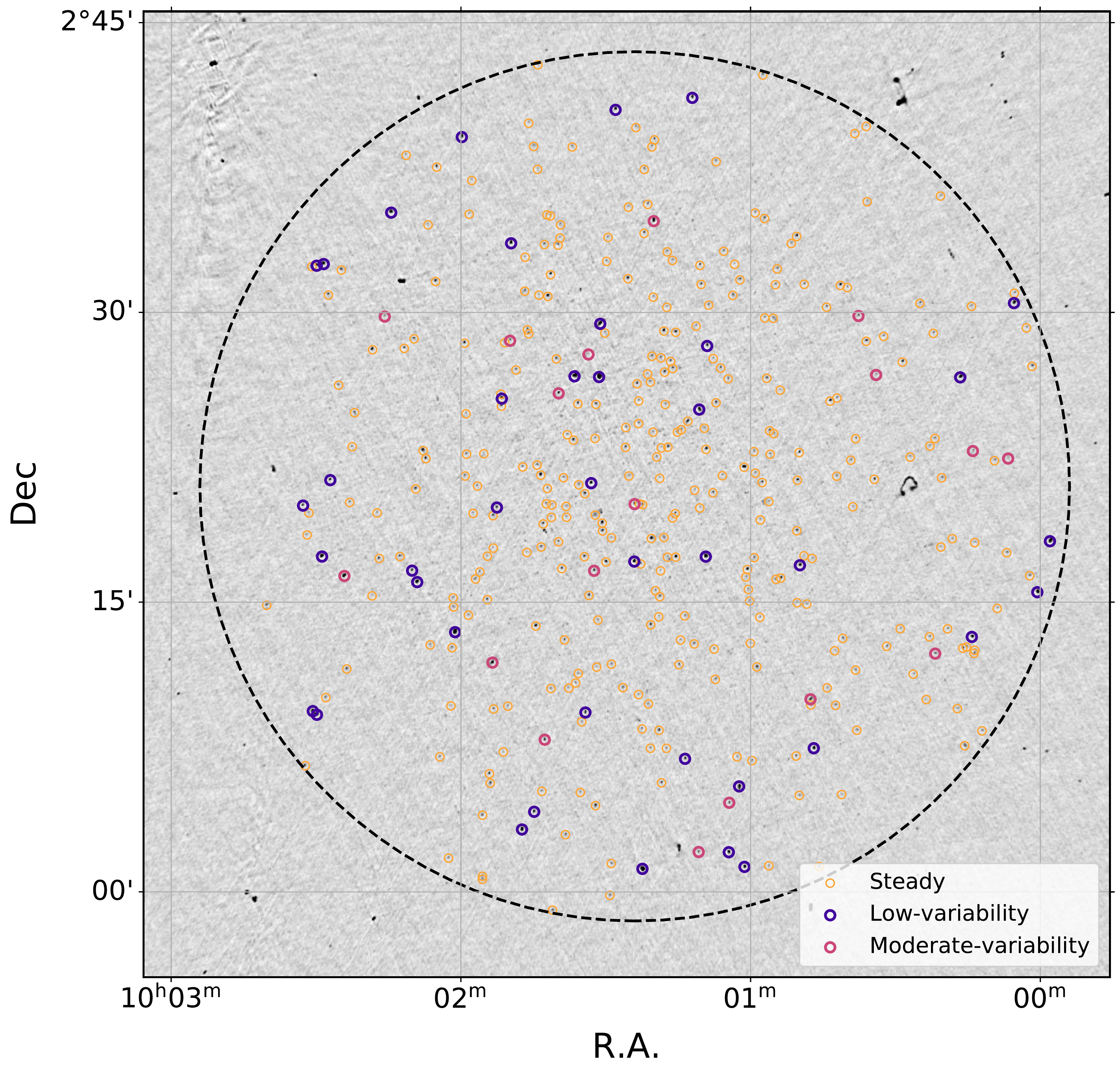}
    \caption{CHILES field showing the 370 variable candidates in our paper (\S \ref{sec:select}) from one of our images from 2017 Nov 16. The large dashed circle has a radius of 22.5$^{\prime}$, and roughly 1.5 times the primary beam HWHM at 1.5 GHz. Red, blue and yellow circles are point-like radio sources identified by TraP that were designated moderate-variability, low-variability and steady sources respectively, as described in \S \ref{sec:varstats}.}
    \label{fig:chiles}
\end{figure*}
Our selection criteria in the previous section yielded a catalog of 370 sources inside the CHILES primary beam (Figure \ref{fig:chiles}) from which we search for variables. As described in \S\ref{sec:select}, we selected for sources that appear compact in our images, so we will henceforth refer to this sample of 370 sources as `point-like radio sources'. Based on $V$ and $\eta$ (\S \ref{sec:identifyvars}), these sources are classified as either `moderate-variability' (showing significant, moderate-amplitude variability), `low-variability' sources (showing significant variability, but of relatively low amplitude), and non-variables (showing no significant variability; in this paper, we will call them `steady' sources).

\begin{figure}
\centering
    \includegraphics[width=0.5\textwidth]{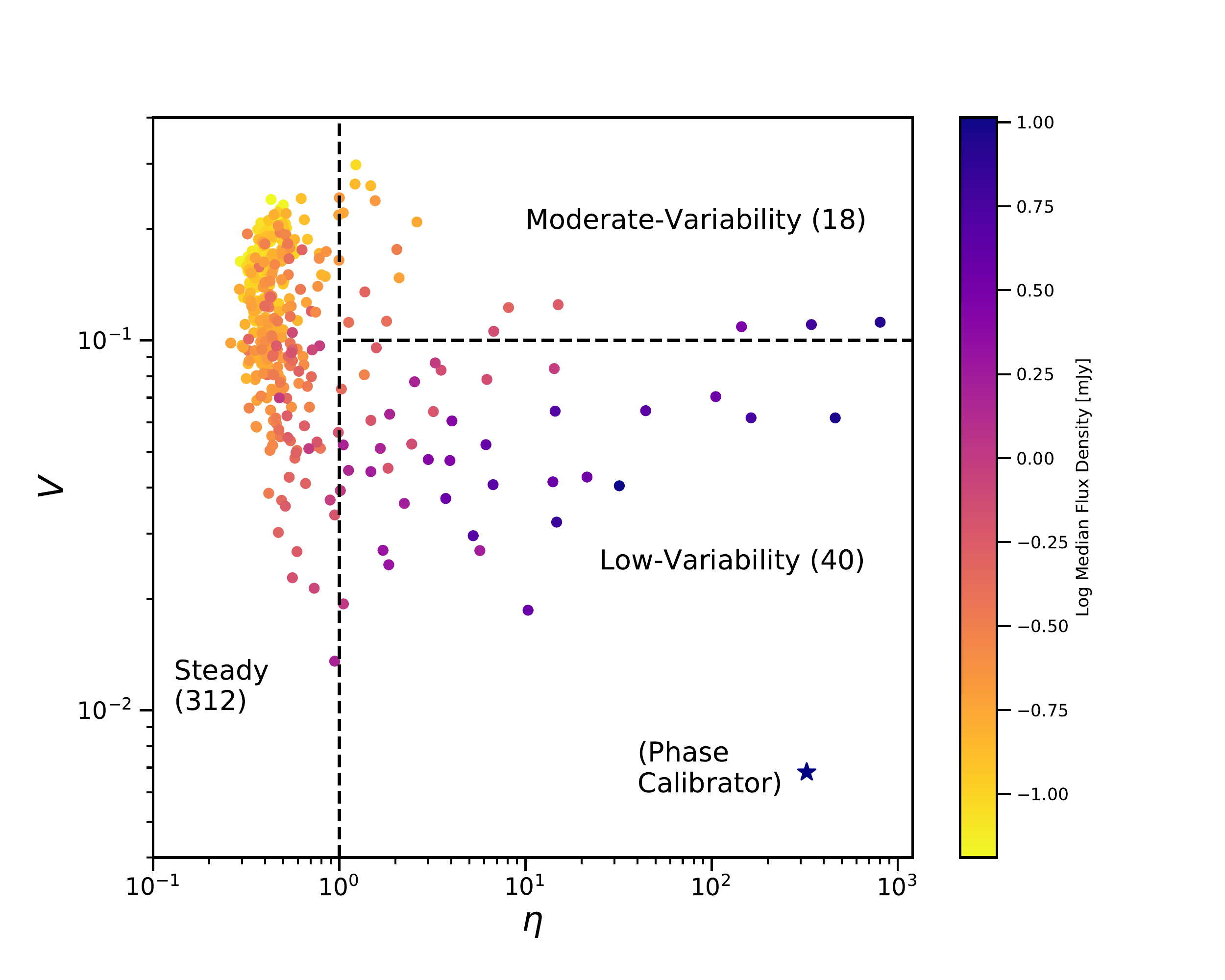}
    \caption{V-$\eta$ distribution of the 370 point-like sources, from which we identify variables, color-coded by their median flux density. Sources at the left of the plot, with $\eta \lesssim 1$, do not have statistically significant variability and are designated `steady'. Higher $\eta$ sources are classified as `moderate-variability' and `low-variability', differentiated at $V=0.1$. The blue star shows the V-$\eta$ position of the complex gain (phase) calibrator, J0943$-$0819. }
\label{fig:Veta}
\end{figure}

Figure\ \ref{fig:Veta} shows the $V-\eta$ distribution of the 370 point-like radio sources. 
The sample has means of $\bar{V} = 0.115\pm0.055$ and $\bar{\eta}=0.43\pm0.13$. Most sources fall below the threshold $\eta=1$ determined in \S \ref{sec:identifyvars}, as
expected for sources with no statistically significant variation \citep{Rowlinson2016, Stewart2016, Rowlinson2019}; these are called ``steady" sources. We are left with 58 sources with $\eta>1$, which are divided into ``moderate-variabiity" and ``low-variability" categories at the threshold of $V=0.1$, corresponding to 10$\%$ variability in integrated flux density.
We note that this threshold was chosen purely for discussion purposes, to separate sources that seemed to show more variability 
than others; the threshold is not a physical cutoff 
(in fact as we show in Figure \ref{fig:rlcsflowvars}, some of the low-variability sources have light curves that are quite similar to moderate-variability ones). Out of the 58 sources, 18 are classified as moderate-variability, with $V\approx0.1-0.3$ while 40 are classified as low-variability with $V=0.02-0.1$.

\begin{figure}
    \centering
    \includegraphics[width=\columnwidth]{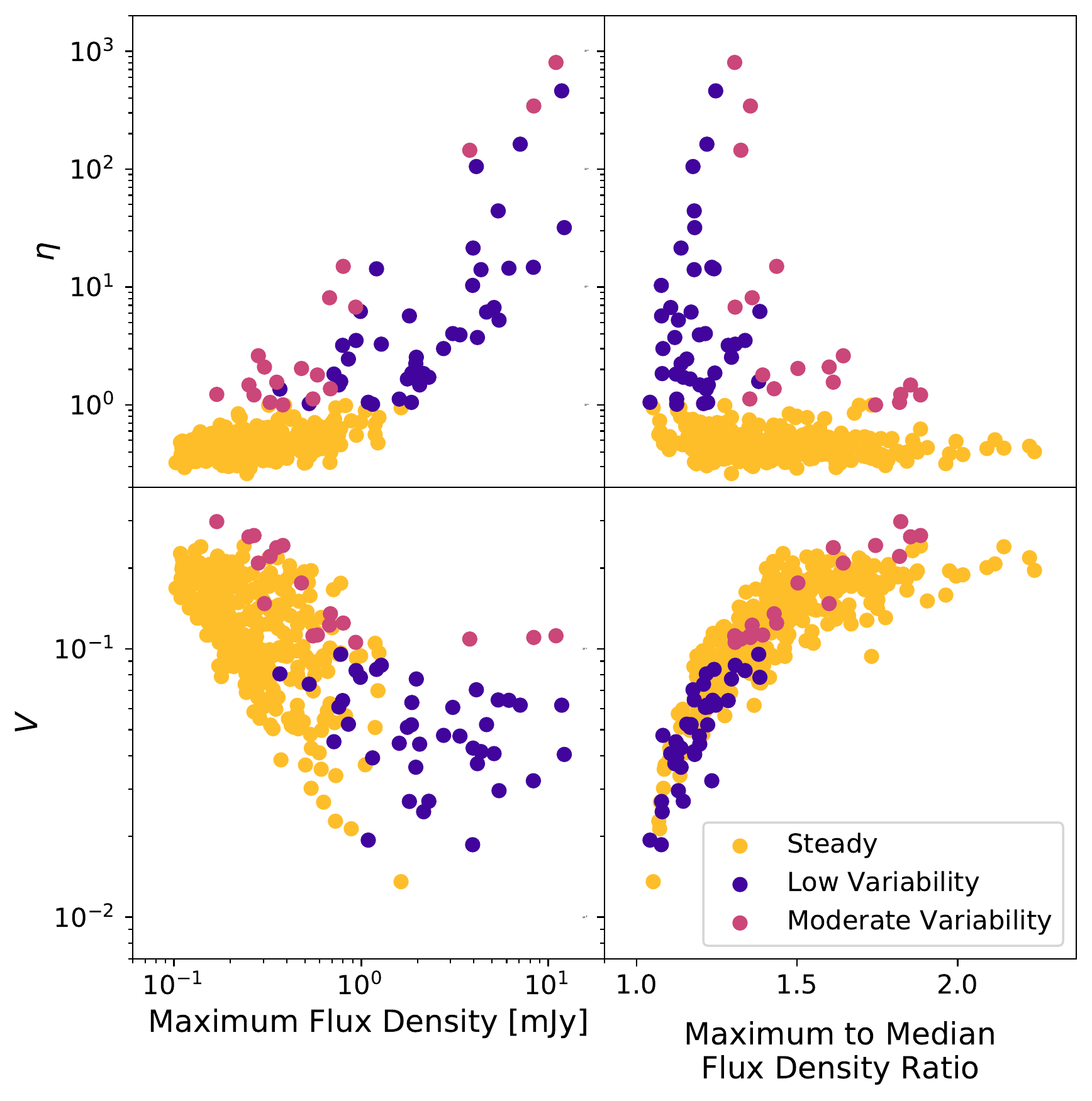}
    \caption{Values of $V$ and $\eta$ plotted against the maximum flux density attained during observations and the ratio of maximum to median flux density. The points are color-coded so  that steady sources are gold, low-variability sources blue, and moderate-variability sources magenta.}
    \label{fig:etaVmaxmed}
\end{figure}

In Figure \ref{fig:etaVmaxmed}, we assess the $V-\eta$ parameter space further,
by comparing $V$ and $\eta$ values with the brightest per-epoch flux density measured in our CHILES VERDES light curves (called maximum flux density) and the ratio of this maximum flux density to the median flux density for that source, measured over its light curve. Not surprisingly (and as also seen in Figure \ref{fig:Veta}), the high-$\eta$ sources are also the brightest sources; a source needs to be detected at high significance to make high-significance measurements of its variability. Variability ($V$) appears to be anti-correlated with flux density, with low-$V$ sources being some of the brightest objects. 
Again, this is not surprising---small-amplitude variations can be securely measured if a source is bright and high-significance. Fainter sources need to have higher $V$ values in order to reach a given $\eta$ value, compared to brighter sources. High $V$, high $\eta$ sources would be easily detectable in our sample, but are not present.

The steady sources 
have flux densities $\lesssim$1.5 mJy, and their $V$ values are inversely correlated with maximum flux density, but positively correlated with max-to-median flux density ratio. Both these trends can be explained if the epoch-by-epoch variability in steady sources is random noise on the order of the image RMS. For example, consider a case where the deviation in flux density from the median, as well as the uncertainty in flux density measurements, are similar to the average RMS of the images ($\sigma_{rms}$); then, $I_i - \xi_{I} \approx \sigma_{rms}$ and $\sigma_i \approx C\sigma_{rms}$, where $C$ is a constant ($\gtrsim 1$ to account for the fact that local RMS might be higher than the image RMS).
In that case, $\eta \lesssim 1$ from Eq.\ \ref{eq:eta}, as observed for steady sources. Similarly for $V$ as defined in Eq.\ \ref{eq:V}, if we assume $s \approx \sigma_{rms}$ and $I_{max} \approx (1-2)\xi_{I}$, then $\bar{V} \propto I_{max}^{-1}$. In other words, steady sources whose fluctuations are mainly driven by the RMS noise of images will have lower $V$ for brighter (and therefore higher signal-to-noise detected) objects. 
The moderate- and low-variability sources on the other hand are as bright as $\sim 10$ mJy,
implying that, contrary to the steady sources, both moderate and low-variability sources undergo flux density changes that are significantly higher than their measurement uncertainties (a fact also reflected in the SFs shown later).

The bottom panels of Figure \ref{fig:etaVmaxmed} shows that the sources span almost two orders of magnitude in maximum flux density (0.1--10 mJy), but their maximum flux density is within a factor of 2 of their average brightness, consistent with previous deep-field variability surveys \citep[e.g.][]{Mooley2013, Bhandari2018, Radcliffe2019}. 
Moderate-variability sources show a higher max-to-median flux density ratio than low-variability sources (they undergo up to a factor of 2 in brightening), explaining their higher $V$. Unlike steady sources, both moderate- and low-variability sources show $V$ values that are uncorrelated with flux density, at least for maximum flux densities above 1 mJy, further supporting that their variability is distinct from noise fluctuations in images. Fainter variable sources however show some correlation between $V$ and maximum flux density like the steady sources, and this is likely due to their lower signal-to-noise detections which makes them more vulnerable to image noise. 

\subsection{Individual Light Curves}
\begin{figure*}
    \centering
    \includegraphics[width=\textwidth]{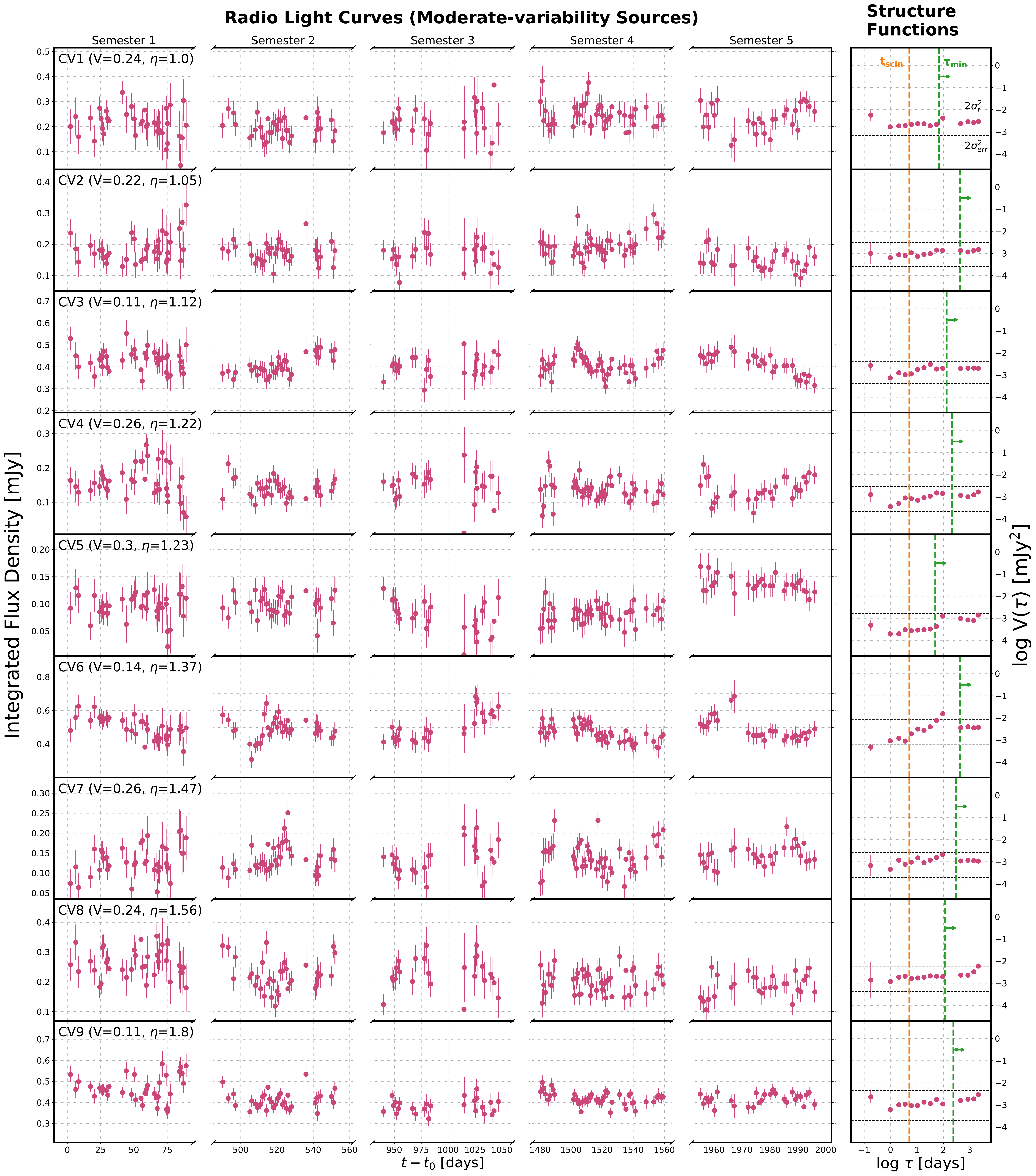}
    \caption{Light curves and structure functions for sources designated `moderate-variability' (\S \ref{sec:varstats}), in increasing order of $\eta$. Each light curve (left panel) shows the integrated flux density and their uncertainties over our five observing semesters, as measured by TraP per CHILES epoch in units of days starting from our first observation, $t_0 = 2456588.5$ JD (2013 Oct 23). The right panel shows the structure function, calculated as described in \S \ref{sec:sf} from the light curve. The upper and lower horizontal black dashed lines show the variance of the flux densities (2$\sigma_f^2$) and variance of the flux density measurement uncertainties (2$\sigma_{err}^2$) respectively. The vertical orange dashed line shows the estimated scintillation timescale along the CHILES line of sight, and the green dashed line shows the minimum variability timescale needed to not exceed the inverse Compton limit on brightness temperatures (see \S \ref{sec:sourceofvars}).}
    \label{fig:rlcsfvars1}
\end{figure*}

\begin{figure*}
    \centering
    \includegraphics[width=\textwidth]{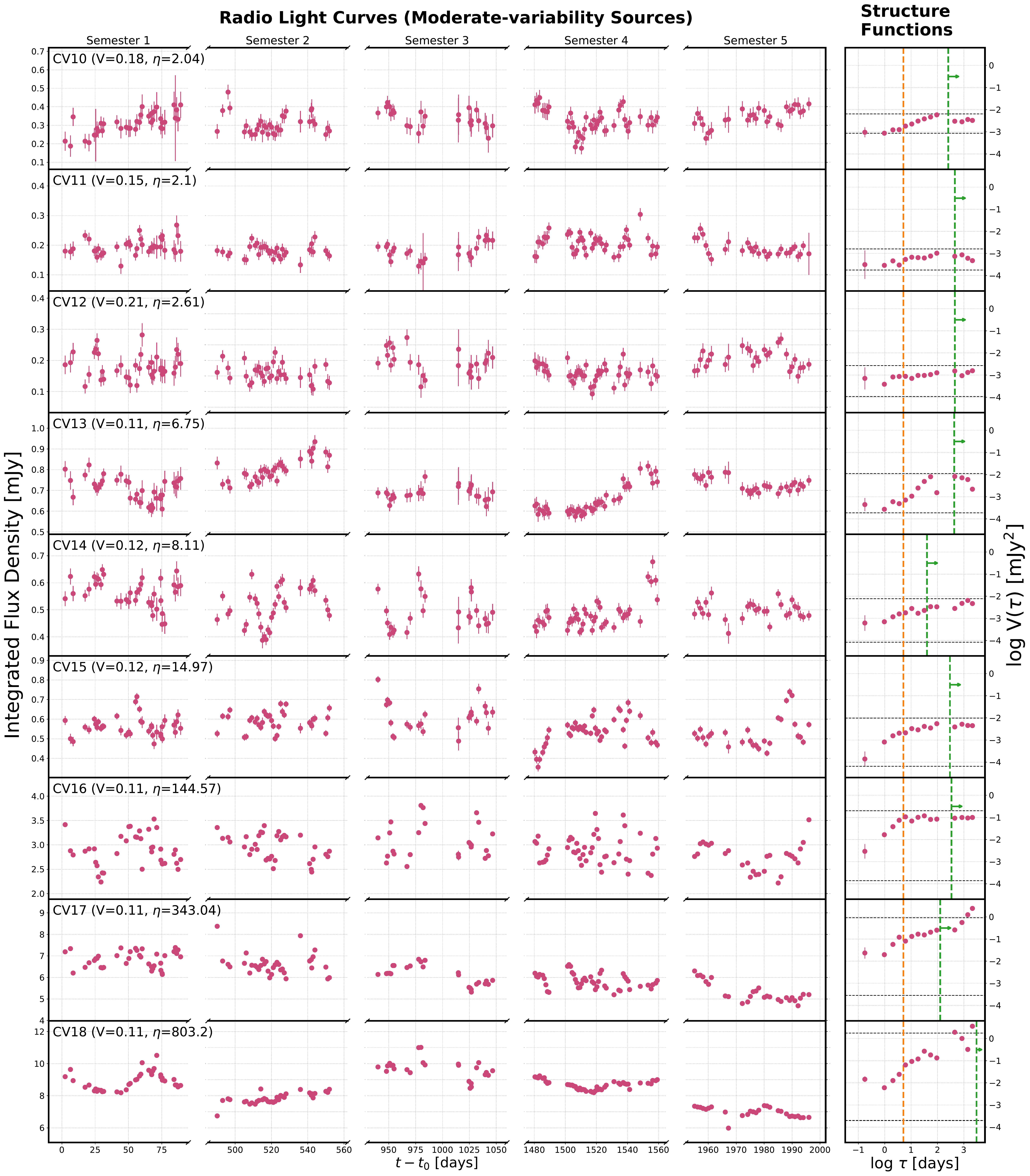}
    \caption{Light Curves (continued), as in Figure \ref{fig:rlcsfvars1}. }
    \label{fig:rlcsfvars2}
\end{figure*}

\begin{figure*}
    \centering
    \includegraphics[width=\textwidth]{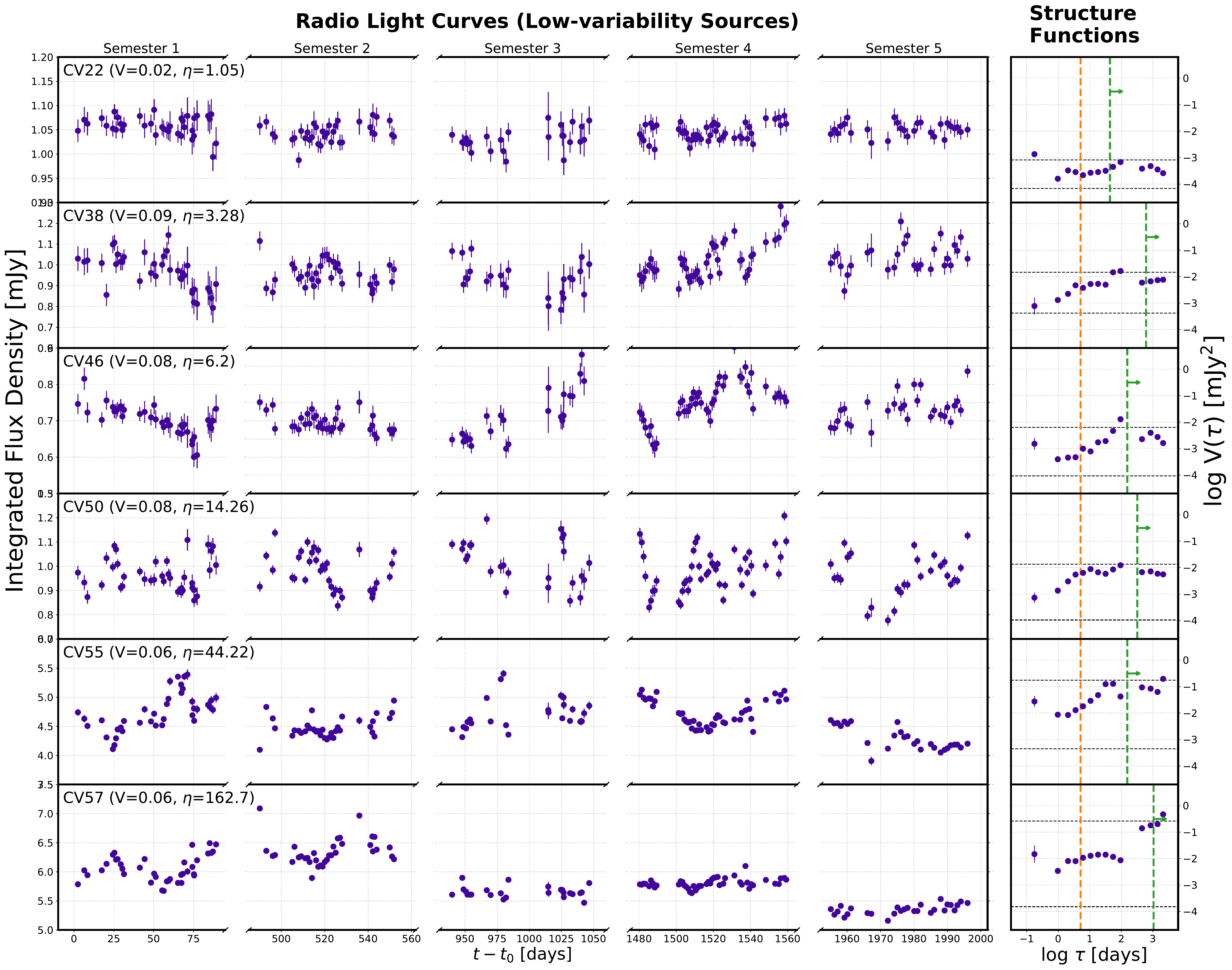}
    \caption{Light Curves of 6 out of the 40 low-variability sources in our sample. The first two light curves are representative of the majority of the sample (most light curves look like these). The last four are sources that show interesting variability by eye, even though they do not pass the selection cut for being ``variable''. Plots are as described as in Figure \ref{fig:rlcsfvars1}, and again $t_0 = 2456588.5$ JD.}
    \label{fig:rlcsflowvars}
\end{figure*}

In this section, we review basic features of the individual light curves and SFs of our moderate-variability and low-variability sources shown in Figures \ref{fig:rlcsfvars1}, \ref{fig:rlcsfvars2} and \ref{fig:rlcsflowvars}. We only show 6 out of the 40 low-variability sources out of consideration for space (the full light curve tables will be available in electronic form), although the majority of them are similar to the first 2 sources in Figure \ref{fig:rlcsflowvars}. From hereon, we name our CHILES VERDES variables with the prefix `CV'. Moderate-variability sources are numbered from 1--18 in order of increasing $\eta$, and low-variability sources are numbered from 19--58, also in order of $\eta$.

The variability patterns in the radio light curves are quite complex, with various episodes of brightening, dimming and jittering on a variety of timescales ranging from days to years. The majority (15 of the 18) of moderate-variability sources are fainter than 1 mJy in average brightness, with variability on the order of 10-30$\%$ over the full course of CHILES observations. Evidence of monthly variability is seen clearly in cases such as CV13, as well as CV10 and CV6. More jittery short timescale variability however is seen in cases like CV14. CV7 shows a rare example of a brightening event by more than a factor of 2 (in semester 2 between $t-t_0$ = 520--540 days) that is well-resolved by our observing cadence (i.e. longer than the timescale between subsequent epochs, but much shorter than a typical semester). Only three moderate-variability sources---CV16, CV17 and CV18--are brighter than 1 mJy, but are among the most prominent moderate-variability in our sample. Among them is CV18, which as we show later, is also the most luminous and distant radio variable source in our sample. CV18 shows variability on monthly timescales (e.g., the fluctuations in the first semester), but also a brightening and dimming on a longer several-year timescale as seen across the rest of the semesters. CV17 shows shorter term variability at the level of $\sim5\%$ on timescales of a few days compared to CV18, but also a systematic dimming by about 2 mJy (or $26\%$) over the full CHILES observation time. The SF does not plateau on the longest timescales, indicating that any characteristic timescale driving the long-term variability must be $>6$ years. CV16 is the other $>$mJy source, and shows mostly variability on timescales of about a week (11.5$\%$ in flux density), without any discernible long-term variability as observed in the other two cases.


\begin{figure*}
    \includegraphics[width=\textwidth]{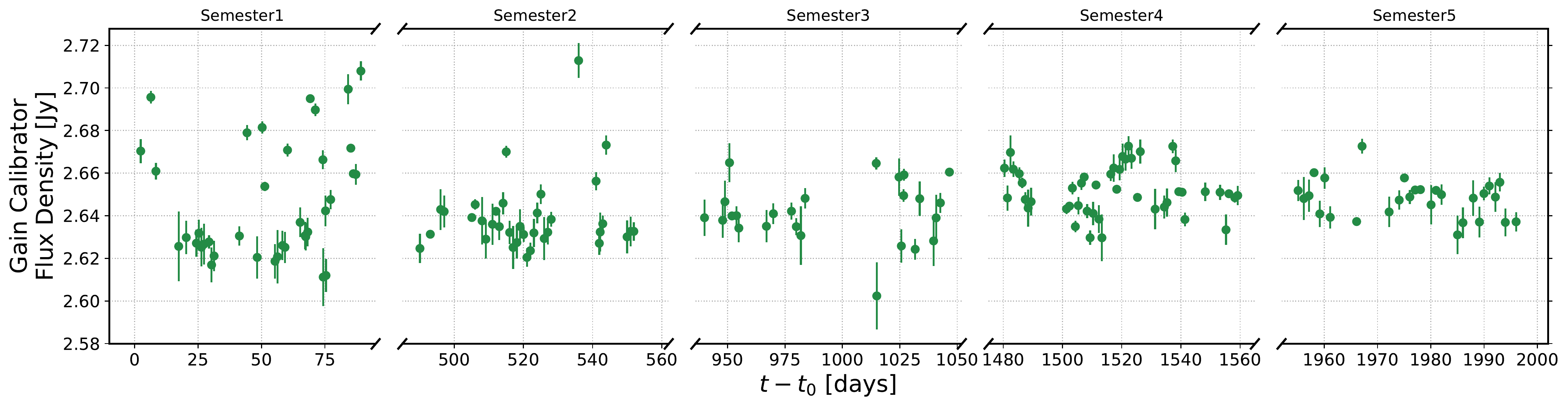}
    \caption{Flux density of the gain-calibrator J0943-0819 measured using the CASA (version 4.7.2) task \texttt{fluxscale} at the end of calibrations.}
    \label{fig:calflux}
\end{figure*}

The majority of the low-variability sources, true to their classification, have low-amplitude fluctuations at the few percent level as evidenced by their light curves and SFs  (e.g. CV22, Figure \ref{fig:rlcsflowvars}). Aside from their low-amplitude fluctuations, their variability characteristics are not all that dissimilar from the moderate-variability sources. For example, CV55 and CV57 are similar to CV18, one of the prominent variables discussed in the previous paragraph. CV55 shows monthly variability seen prominently in Semesters 1 and 4, but also a slower variability across the semesters on a rough timescale of 5.4 years based on the SF. CV57 similarly shows both short and long timescale variability. More short-term jittering variation in flux density is seen in CV50's light curve, with about 8$\%$ variation in flux density occurring on timescale of about 10 days. CV46 is an example of a quiescent source that underwent a sudden short brightening. The source had a median flux density of 0.67 mJy and only 5$\%$ flux density variation until the middle of semester 3, then brightened by about 44$\%$ to 1 mJy in about 2.5 weeks. It is unknown how far the source actually brightened since we do not have reliable observations afterwards. 

In summary, we find that the moderate-variability (and some low-variability sources) show a range of time-series behavior, with variations on both short and slow timescales, and the occasional brightening during a quiescent phase. 

\subsection{Variability due to calibration} 
One pertinent question is the amount of observed variability induced by the calibration process of CHILES VERDES data. Flux densities in each CHILES epoch depend on the quality of calibration solutions 
as well as any intrinsic variability in the flux calibrator 3C286
. We can get an idea of these effects from changes in the flux-density scale of the complex gain calibrator J0943$-$0819, whose flux density is measured in CHILES VERDES epoch by comparison with 3C286, 
shown in Figure \ref{fig:calflux}. J0943$-$0819 shows a low-level scatter in its flux density values of roughly $0.2-0.3\%$ per semester, except the first semester where it shows a slightly elevated level of variability ($\sim 0.6\%$). There is also a hint of a slow rise of $0.6\%$ in median flux density over 5.5 years. While we cannot rule out the variability of J0943-0819 itself, we note that this calibrator's flux destiny scale is set by 3C 286. Therefore, any very low level variability exhibited by 3C286 could have been transferred to J0943-0819. 
A similar scatter is also observed in the polarization time-series of J0943$-$0819, which was also attributed to variability in 3C286 \citep{Hales2019}. Such low-level  ($<1\%$) variability in 3C286 is consistent with archival measurements \cite{Perley2013a, Perley2013b, Perley2017}.

\begin{figure*}
    \includegraphics[width=\textwidth]{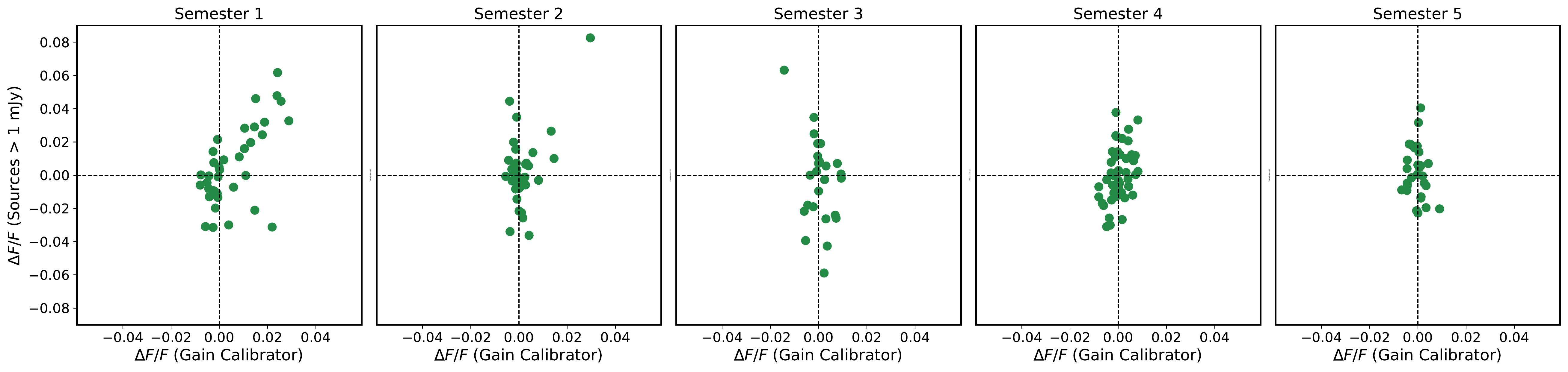}
    \caption{Comparison of the fractional deviation from median flux density ($\Delta F/F$) for the gain calibrator J0943-0819 and the sum of all sources brighter than 1 mJy.}
    \label{fig:caldeltaflux}
\end{figure*}

The variability in the calibrators will likely have a relatively small contribution to the variability observed in our targets. This is particularly evident from the $V-\eta$ values of J0943$-$0819 (blue star in Figure \ref{fig:Veta}), compared to the CHILES VERDES sources. J0943$-$0819 has a high $\eta$ because it is bright (2.7 Jy at 1.45 GHz), observed at high signal-to-noise, but it has $V \approx 0.007$ calculated over the full CHILES VERDES observations---more than an order of magnitude lower than $V$ of CHILES targets. We also compare the fractional deviation in flux density from the median flux density per epoch ($\Delta F/F$) for J0943$-$0819, and for the summed flux density of all sources brighter than 1 mJy. We define the quantity as:
\begin{equation}
    \Delta F/F = \frac{F_i - \langle F \rangle}{\langle F \rangle}
    \label{eq:deltaF}
\end{equation}
where $F_i$ is the flux density in an epoch $i$ (of the calibrator or the mJy-bright CHILES sources) and $\langle F \rangle$ is the median flux density during a particular CHILES semester. Figure \ref{fig:caldeltaflux} shows that the scatter in $\Delta F/F$ for the CHILES VERDES sources is larger than the gain calibrator (consistent with Figure \ref{fig:Veta}) and mostly uncorrelated with the gain calibrator variability, An exception is the first semester, when some anomalously high flux densities of the calibrator (also seen in Figure \ref{fig:calflux}) are correlated with the variability of the CHILES sources. 
However, even if we exclude the 2013 semester, the impact on the $V-\eta$ distribution in Figure \ref{fig:Veta} is minimal, with an average difference of $6\%$ in the $V-\eta$ measurements with and without the 2013 semester. We conclude therefore that statistically speaking, issues with calibration are unlikely to be producing the observed variability in CHILES VERDES.

\begin{deluxetable*}{lccccrccrlr} \label{tab:variables}
\tablecaption{Multi-band and multi-wavelength properties of our 18 CHILES VERDES moderate-variability sources. The column headers are as follows: ``ID'': CHILES VERDES ID assigned to the object, ``R.A.,Dec'': ICRS Right Ascension and Declination found by TraP, ``$\langle F \rangle$'': Median 1.4 GHz flux density measured by CHILES VERDES for all 172 epochs, ``$V, \eta$'': Coefficient of variability and reduced $\chi^2$ statistic defined in \S \ref{sec:varstats}, ``Log\ L$_{1.4}^{S10}$'': VLA COSMOS 1.4 GHz luminosity from \cite{Smolcic2017b}, ``Log L$_X^{M16}$'': 0.5--8 keV X-ray luminosity from \cite{Marchesi2016}, ``$\alpha$'': 1.4--3 GHz spectral index, ``$z$'': redshift, and ``Classification'': Classification of the objects as HLAGN, MLAGN or Star-forming galaxy (SFG) as discussed in \S \ref{sec:multiwav}.}
\tablehead{\colhead{ID} & \colhead{R.A.} & \colhead{Dec} & \colhead{$\langle F \rangle$} & \colhead{$V$} & \colhead{$\eta$} & \colhead{Log L$_{1.4}^{S10}$} & \colhead{Log L$_X^{M16}$} & \colhead{$\alpha$} & \colhead{$z$} & \colhead{Classification}\\
& & & (mJy) & & & (W/Hz) & (ergs/s) & & & }
\startdata
CV1 & 10:00:06.6 & +02:22:26.1 & 0.22 & 0.24 & 1.00 & 22.17 & $<$41.24 & 0.19 $\pm$ 0.47 & 0.222 & -- \\
CV2 & 10:02:15.8 & +02:29:47.0 & 0.18 & 0.22 & 1.05 & 23.58 & $<$42.90 & 0.65 $\pm$ 0.43 & 1.146 & MLAGN\\
CV3 & 10:00:21.8 & +02:12:20.2 & 0.41 & 0.11 & 1.12 & 23.37 & $<$41.89 & -0.49 $\pm$ 0.22 & 0.426 & MLAGN\\
CV4 & 10:00:37.6 & +02:29:49.0 & 0.14 & 0.27 & 1.22 & 23.33 & 42.3 & 0.06 $\pm$ 0.5 & 0.671 & HLAGN\\
CV5 & 10:01:33.6 & +02:27:49.5 & 0.09 & 0.30 & 1.23 & 22.22 & $<$41.21 & -0.83 $\pm$ 0.58 & 0.215 & MLAGN\\
CV6 & 10:01:10.8 & +02:02:04.1 & 0.48 & 0.14 & 1.37 & 23.84 & 42.87 & 0.99 $\pm$ 0.27 & 0.972 & HLAGN\\
CV7 & 10:00:34.0 & +02:26:45.7 & 0.14 & 0.26 & 1.47 & 23.49 & $<$42.65 & -0.02 $\pm$ 0.51 & 0.9 & MLAGN\\
CV8 & 10:00:13.9 & +02:22:49.4 & 0.22 & 0.24 & 1.56 & 22.63 & 42.25 & 0.73 $\pm$ 0.46 & 0.347 & HLAGN\\
CV9 & 10:00:47.6 & +02:09:58.5 & 0.42 & 0.11 & 1.80 & 24.31 & $<$42.34 & -1.19 $\pm$ 0.21 & 0.669 & MLAGN\\
CV10 & 10:01:04.4 & +02:04:37.1 & 0.32 & 0.18 & 2.04 & 23.51 & $<$42.34 & -0.14 $\pm$ 0.34 & 0.668 & MLAGN\\
CV11 & 10:01:32.4 & +02:16:37.6 & 0.19 & 0.15 & 2.10 & 24.08 & $<$43.08 & -0.34 $\pm$ 0.29 & 1.364 & MLAGN\\
CV12 & 10:01:49.8 & +02:28:32.0 & 0.17 & 0.21 & 2.61 & 23.76 & $<$42.99 & 0.24 $\pm$ 0.4 & 1.248 & MLAGN\\
CV13 & 10:01:42.6 & +02:07:52.9 & 0.71 & 0.11 & 6.75 & 24.39 & $<$42.64 & -0.51 $\pm$ 0.21 & 0.89 & MLAGN\\
CV14 & 10:01:39.8 & +02:25:48.6 & 0.5 & 0.12 & 8.11 & 22.34 & 41.79 & -0.398 $\pm$ 0.244 & 0.124 & MLAGN\\
CV15 & 10:01:24.0 & +02:20:04.7 & 0.56 & 0.13 & 14.97 & 23.92 & $<$42.34 & -0.406 $\pm$ 0.248 & 0.666 & MLAGN\\
CV16 & 10:01:53.5 & +02:11:52.4 & 2.87 & 0.11 & 144.57 & 23.07 & 43.01 & -0.28 $\pm$ 0.21 & 0.405 & HLAGN \\
CV17 & 10:02:24.1 & +02:16:21.3 & 6.18 & 0.11 & 343.04 & 23.55 & 41.13 & -0.67 $\pm$ 0.21 & 0.122 & -- \\
CV18 & 10:01:20.1 & +02:34:43.6 & 8.43 & 0.11 & 803.20 & 25.77 & 43.77 & 0.097 $\pm$ 0.218 & 1.555 & HLAGN
\enddata
\end{deluxetable*}

\begin{deluxetable*}{lccccrccrlr} \label{tab:lowvariables}
\tablecaption{Multi-band and multi-wavelength properties of our 40 CHILES VERDES low-variability sources. The column headers have the same meaning as in Table \ref{tab:variables}.}
\tablehead{\colhead{ID} & \colhead{R.A.} & \colhead{Dec} & \colhead{$\langle F \rangle$} & \colhead{$V$} & \colhead{$\eta$} & \colhead{Log L$_{1.4}^{S10}$} & \colhead{Log L$_X^{M16}$} & \colhead{$\alpha$} & \colhead{$z$} & \colhead{Classification}\\
& & & (mJy) & & & (W/Hz) & (ergs/s) & & & }
\startdata
CV19 & 10:02:27.1 & +02:21:19.2 & 0.95 & 0.039 & 1.01 & 24.89 & $<$42.95 & $-0.83 \pm 0.11$ & 1.202 & MLAGN\\
CV20 & 10:01:13.6 & +02:06:53.6 & 0.41 & 0.074 & 1.03 & 23.2 & $<$41.61 & $-0.55 \pm 0.15$ & 0.323 & MLAGN\\
CV21 & 10:01:01.3 & +02:01:18.0 & 1.31 & 0.052 & 1.05 & 25.72 & $<$43.41 & $-1.28 \pm 0.12$ & 1.89 & MLAGN\\
CV22 & 10:01:33.0 & +02:21:09.8 & 1.05 & 0.019 & 1.05 & 24.36 & $<$42.39 & $-0.96 \pm 0.08$ & 0.7 & MLAGN\\
CV23 & 10:01:28.0 & +02:40:29.3 & 1.25 & 0.045 & 1.12 & 24.92 & 42.79 & $-0.75 \pm 0.11$ & 1.108 & HLAGN\\
CV24 & 10:01:52.5 & +02:19:54.2 & 0.30 & 0.081 & 1.36 & 22.87 & $<$41.34 & $-0.71 \pm 0.17$ & 0.246 & HLAGN\\
CV25 & 10:01:04.5 & +02:02:03.5 & 1.51 & 0.044 & 1.48 & -- & -- & $-1.09 \pm 0.11$ & -- & -- \\
CV26 & 10:02:10.1 & +02:16:38.0 & 0.60 & 0.061 & 1.48 & 24.28 & $<$42.63 & $-0.53 \pm 0.13$ & 0.879 & MLAGN\\
CV27 & 10:02:32.7 & +02:20:00.0 & 0.51 & 0.095 & 1.58 & 23.86 & $<$42.25 & $-0.34 \pm 0.20$ & 0.607 & MLAGN\\
CV28 & 10:00:14.2 & +02:13:12.1 & 1.34 & 0.051 & 1.66 & 25.35 & 43.63 & $-1.80 \pm 0.12$ & 1.14 & HLAGN\\
CV29 & 10:00:46.9 & +02:07:26.5 & 1.85 & 0.027 & 1.72 & 25.31 & $<$42.99 & $-1.02 \pm 0.08$ & 1.252 & HLAGN\\
CV30 & 10:01:10.6 & +02:24:58.4 & 0.63 & 0.045 & 1.83 & 25.71 & $<$43.30 & $-1.83 \pm 0.11$ & 1.698 & MLAGN\\
CV31 & 10:01:49.6 & +02:33:34.8 & 1.89 & 0.025 & 1.84 & 24.88 & $<$42.52 & $-1.29 \pm 0.08$ & 0.792 & MLAGN\\
CV32 & 10:02:28.4 & +02:32:30.2 & 1.31 & 0.063 & 1.86 & -- & -- & -- & -- & -- \\
CV33 & 10:01:02.4 & +02:05:27.9 & 1.58 & 0.036 & 2.23 & -- & -- & -- & -- & -- \\
CV34 & 10:01:34.2 & +02:09:17.5 & 0.71 & 0.052 & 2.46 & 25.32 & $<$43.57 & $-0.76 \pm 0.12$ & 2.222 & MLAGN\\
CV35 & 10:02:30.7 & +02:09:21.6 & 1.32 & 0.077 & 2.54 & -- & -- & -- & -- & -- \\
CV36 & 10:01:12.1 & +02:41:06.6 & 2.21 & 0.048 & 3.01 & 25.46 & 43.12 & $-1.30 \pm 0.11$ & 1.258 & HLAGN\\
CV37 & 10:01:09.0 & +02:28:15.8 & 0.61 & 0.064 & 3.2 & 24.32 & $<$42.66 & $-0.57 \pm 0.15$ & 0.911 & MLAGN\\
CV38 & 10:01:44.8 & +02:04:09.1 & 0.89 & 0.087 & 3.28 & 24.5 & $<$42.72 & $-0.43 \pm 0.18$ & 0.965 & MLAGN\\
CV39 & 10:01:36.5 & +02:26:41.6 & 0.69 & 0.083 & 3.52 & 22.97 & 41.06 & $-1.25 \pm 0.15$ & 0.123 & MLAGN\\
CV40 & 10:01:59.8 & +02:39:04.8 & 3.24 & 0.037 & 3.73 & 25.04 & $<$42.53 & $-0.79 \pm 0.10$ & 0.8 & MLAGN\\
CV41 & 10:02:29.9 & +02:32:25.1 & 2.47 & 0.047 & 3.93 & 24.55 & 43.37 & $-0.76 \pm 0.09$ & 0.432 & HLAGN\\
CV42 & 10:02:29.8 & +02:09:10.3 & 2.21 & 0.061 & 4.03 & -- & -- & -- & -- & -- \\
CV43 & 10:02:14.5 & +02:35:10.0 & 4.28 & 0.03 & 5.24 & 26.07 & $<$43.51 & $-0.67 \pm 0.09$ & 2.099 & MLAGN\\
CV44 & 10:01:24.1 & +02:17:06.3 & 1.67 & 0.027 & 5.69 & 25.4 & $<$43.29 & $-0.70 \pm 0.09$ & 1.677 & MLAGN\\
CV45 & 10:00:05.4 & +02:30:29.0 & 3.36 & 0.052 & 6.13 & 24.96 & $<$42.46 & $-0.77 \pm 0.12$ & 0.746 & MLAGN\\
CV46 & 10:01:51.5 & +02:25:32.2 & 0.70 & 0.078 & 6.20 & 24.22 & 42.55 & $-0.65 \pm 0.16$ & 0.827 & HLAGN\\
CV47 & 10:00:00.6 & +02:15:31.0 & 3.91 & 0.041 & 6.7 & 25.68 & 44.47 & $0.33 \pm 0.10$ & 2.45 & HLAGN\\
CV48 & 10:01:09.3 & +02:17:21.6 & 3.64 & 0.019 & 10.32 & 26.35 & $<$43.72 & $-0.98 \pm 0.07$ & 2.582 & MLAGN\\
CV49 & 10:02:28.8 & +02:17:21.8 & 3.41 & 0.041 & 14.03 & 26.07 & 43.68 & $-0.57 \pm 0.10$ & 2.625 & HLAGN\\
CV50 & 10:00:49.8 & +02:16:54.8 & 0.95 & 0.084 & 14.27 & 24.53 & $<$42.67 & $-0.53 \pm 0.17$ & 0.917 & MLAGN\\
CV51 & 09:59:58.0 & +02:18:09.3 & 4.20 & 0.064 & 14.43 & 26.18 & $<$43.30 & $-1.02 \pm 0.14$ & 1.698 & MLAGN\\
CV52 & 10:01:47.3 & +02:03:14.1 & 6.01 & 0.032 & 14.70 & 24.36 & 42.16 & $-0.74 \pm 0.09$ & 0.323 & HLAGN\\
CV53 & 10:02:01.2 & +02:13:27.0 & 3.35 & 0.043 & 21.40 & -- & -- & $-0.99 \pm 0.10$ & -- & -- \\
CV54 & 10:01:22.4 & +02:01:11.9 & 9.04 & 0.04 & 31.90 & 25 & $<$41.88 & -- & 0.425 & HLAGN\\
CV55 & 10:00:16.6 & +02:26:38.3 & 4.15 & 0.065 & 44.22 & 26.3 & $<$43.66 & $-0.77 \pm 0.14$ & 2.436 & HLAGN\\
CV56 & 10:01:31.1 & +02:29:24.7 & 3.44 & 0.07 & 105.14 & 24.31 & 42.81 & $-0.78 \pm 0.15$ & 0.349 & HLAGN\\
CV57 & 10:02:09.1 & +02:16:02.4 & 5.58 & 0.062 & 162.70 & 25.34 & $<$42.68 & $-0.71 \pm 0.13$ & 0.928 & -- \\
CV58 & 10:01:31.4 & +02:26:39.5 & 9.39 & 0.062 & 460.92 & 24.79 & $<$41.68 & $-0.76 \pm 0.13$ & 0.348 & MLAGN
\enddata
\end{deluxetable*}

\section{multi-wavelength Characteristics} \label{sec:multiwav}
\subsection{AGN classification} \label{sec:s17class}
We can determine the physical characteristics of our radio-variable sources using the plethora of multi wavelength data available in the COSMOS field\footnote{\url{http://cosmos.astro.caltech.edu/page/datasets}}. A large number of surveys have targeted the COSMOS field for characterizing galaxies, but in this paper we will closely follow the work of \citet[][henceforth S17]{Smolcic2017b}, who provide the largest catalog of $z<6$ \emph{radio-selected} galaxies classified using observations spanning radio to X-ray wavelengths. S17 uses the VLA-COSMOS 3 GHz survey, which detected 10,830 sources above 5$\sigma$ with 384 hrs of VLA data in the full 2.2 sq deg of the COSMOS field \citep{Smolcic2017a}.\ S17 obtained 8035 multi-wavelength counterparts for the 3 GHz sources in the unmasked areas of the COSMOS field  (masked areas mainly demarcate regions with bright foreground stars, gaps between chips, etc. in data at non-radio wavelengths). 

Radio sources in extragalactic surveys are mainly star-forming galaxies and AGN \citep{Condon1992}. Radio emission from star-forming galaxies mainly traces the integrated cosmic-ray population produced by supernovae and supernova remnants in massive star-forming regions \citep{Murphy2009, Delhaize2017}.
For AGN, radio emission mainly traces the synchrotron emission produced by the collimated relativistic jet from the central super-massive black hole (SMBH), although outflows and accretion disk coronae can also contribute to radio emission in radio-quiet systems \citep{Panessa2019}. Such radio-selected AGN have been differentiated into two main categories based on the appearance of excitation lines in optical spectra: high-excitation and low-excitation \citep{Smolcic2009, Heckman2014}. The two categories are believed to trace different phases of AGN--galaxy coevolution, with high-excitation sources emitting across
a wide range of the electromagnetic spectrum, and consistent with radiatively efficient SMBH accretion at high bolometric luminosity; these are often associated with galaxies in the ``green valley" of the color-magnitude diagram. 
Meanwhile, low-excitation sources exhibit lower bolometric luminosities as expected for radiatively inefficient SMBH accretion, and tend to be associated with the red, passive-sequence of galaxies \citep{Smolcic2009, Heckman2014}.

S17 expands on this classification further with the abundant multi wavelength information available in the COSMOS field, including the latest ($z^{++}YJHK_S$) photometry from COSMOS2015 \citep{Laigle2016}, the $i$-band catalog from \cite{Capak2007}, the 3.6 $\mu$m Spitzer(IRAC)-COSMOS (S-COSMOS) catalog from \cite{Sanders2007}, FIR $Herschel$ data \citep{Poglitsch2010, Griffin2010, Oliver2012}, sub-mm photometry from various observational campaigns \citep{Scott2008, Casey2013, Aretxaga2011, Bertoldi2007, Smolcic2012, Miettinen2015}, and the X-ray catalog of point sources from the \emph{Chandra}-COSMOS Legacy Survey \citep{Marchesi2016}. These data sets were used to classify the 3-GHz radio sources into three broad categories: high-to-moderate luminosity AGN (HLAGN), moderate-to-low luminosity AGN (MLAGN) and star-forming galaxies. HLAGN and MLAGN are expected to be high-redshift counterparts to the high- and low-excitation galaxies discussed earlier, and the classification scheme is based on bolometric luminosity which also serves as a proxy for the SMBH accretion rate (Figure 9 in S17). We refer the reader to S17 and \citet{Delvecchio2017} for the details of the selection criteria and individual categories, and provide only a brief description below of each category. 

VLA-COSMOS 3 GHz sources were classified as \textbf{HLAGN} if they satisfied at least one of the three criteria -- (i)  0.5--8 keV X-ray luminosity $> 10^{42}$ ergs s$^{-1}$ \citep{Szokoly2004}, (ii) mid-IR colors consistent with the AGN selection criteria of \cite{Donley2012}, or (iii) a significant AGN contribution to the optical-to-mm spectral energy distribution (SED; \citealt{Delvecchio2017}). The individual criteria are limited in their completeness, but together they are believed to effectively select AGN with high radiative luminosities (Figure 9 of S17). 

The remaining sources not classified as HLAGN were grouped into MLAGN and star-forming galaxies based on their rest-frame colors and presence of `radio-excess', i.e.\ 1.4 GHz luminosity exceeding (by more than 3$\sigma$) the redshift-dependent $L_{1.4}$--SFR relation, where the SFR was determined from the integrated IR (1--1000 $\mu$m) luminosity from SED fitting \citep{Delhaize2017, Delvecchio2017}. A non-HLAGN object was classified as \textbf{Star-forming galaxy} if its dust-corrected rest-frame colors (i.e., either $M_{NUV} - M_{r^+} < 3.5$ mag or $M_{NUV} - M_{r^+} > 3.5$ mag  but with detection in Herschel bands), \emph{and} exhibited no radio-excess. The remaining radio sources were classified as \textbf{MLAGN}, and consist of two sub-categories: (i) galaxies deemed to be red/quiescent (based on their dust-corrected rest-frame colors $M_{NUV} - M_{r^+} > 3.5$ mag and non-detection in \emph{Herschel} bands), and (ii) star-forming galaxies (also based on rest-frame colors) but with a $>3\sigma$ radio-excess.

In addition to AGN classifications, we also discuss the
redshifts and radio spectral indices from the VLA-COSMOS 1.4 GHz and 3 GHz surveys, which are described in the subsequent sections.

\subsection{Crossmatching with S17 using TOPCAT}\label{sec:topcat}
We cross-match our sources with the S17 catalog using TOPCAT, a GUI analysis package for working with large tabular datasets \citep{Taylor2005}. CHILES VERDES sources were matched to their nearest S17 sources within a radius of $1^{\prime\prime}$, which is roughly equal (but rounded-off) to the TraP-based positional errors as discussed in \S \ref{sec:select}. We note that S17 took into account possible astrometric uncertainties and false-match probabilities (e.g., due to chance coincidences) when constructing the catalog of multi-wavelength counterparts for the VLA-COSMOS radio sources. Since the CHILES VERDES and S17 catalogs are both based on VLA observations of a common patch of sky, we do not employ any additional cross-match verification beyond visual checks and simply identify the nearest-neighbors for our sources. 

\begin{figure*}[!t]
    \includegraphics[width=\textwidth]{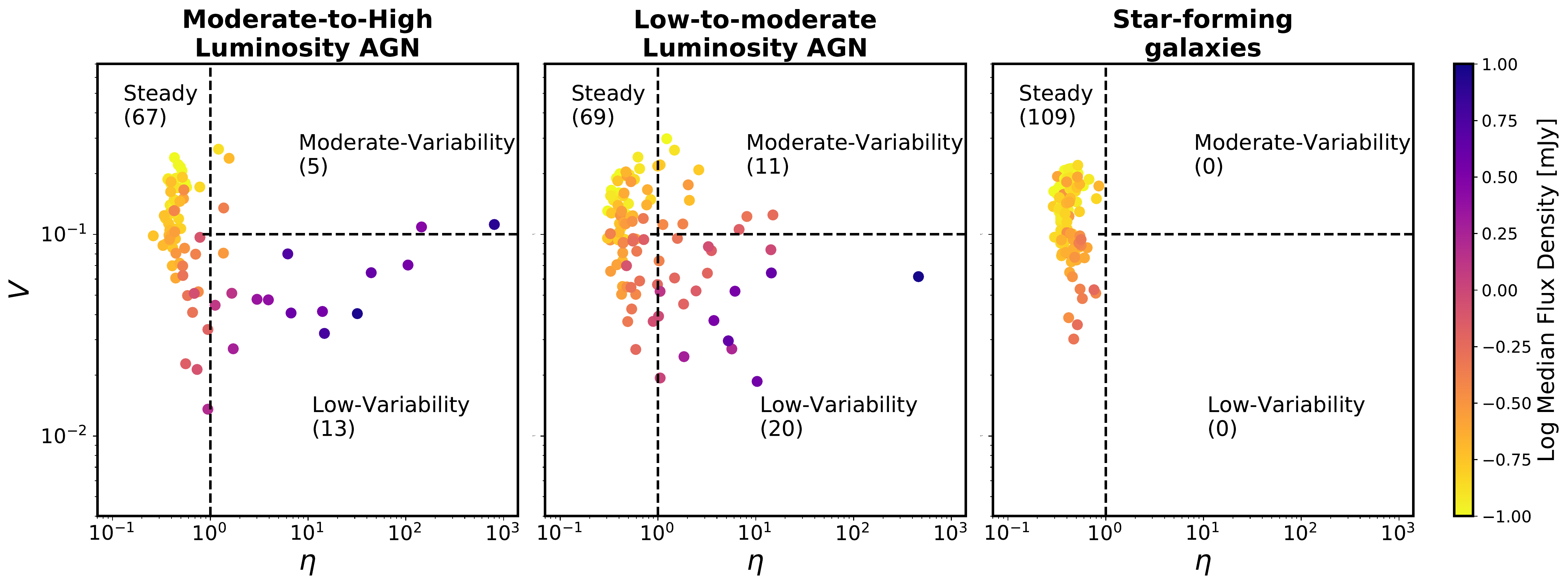}
    \caption{Distribution of CHILES VERDES sources in the $V-\eta$ parameter space, separated by multi-wavelength classification of host galaxies. Each panel has the same `steady--moderate variability--low variability' parameter spaces defined in Figure \ref{fig:Veta}, with colors indicating log 1.4 GHz median flux density. 
    For details on the host galaxy classification, refer to \S \ref{sec:s17class}.}
    \label{fig:multiwav}
\end{figure*}

\subsection{Results}
\subsubsection{Host galaxy properties}
Out of the 370 point-like CHILES VERDES sources, TOPCAT found cross-matches for 316 sources in the \cite{Smolcic2017b} catalog; their demographics are shown in Figure \ref{fig:multiwav}. The remaining 54 sources did not cross-match for several reasons, such as the higher image resolution of the VLA-COSMOS 3 GHz survey (i.e., some unresolved CHILES VERDES sources were resolved into two or more point sources in 3 GHz that were spatially separated on scales larger than our search radius of 1$^{\prime\prime}$), faintness of the 3 GHz counterpart, proximity to a bright, saturated object (thus appearing in a masked region), or a genuine lack of optical/NIR counterparts at the source location. However, the majority of CHILES VERDES point-like sources have counterparts, and form a large enough sample to be sufficiently representative of the different source populations that we discuss below.

Of the variable sources, cross-matches were obtained for 16 out of 18 moderate-variability sources, and 33 out of 40 low-variability sources.
The CHILES VERDES variables fall into both HLAGN and MLAGN categories as seen in Figure \ref{fig:multiwav}, with MLAGN featuring more variables in both moderate and low-variability categories compared to HLAGN. We do note that two of our brightest variables in the sample -- CV16 and CV18 -- are HLAGN. This may suggest that radio variability is correlated with the SMBH accretion rate, although our relatively small sample size makes this a low significance result. 

A notable, if not surprising, result is that all sources that are ``clean" (without radio excess) star-forming galaxies are also ``steady", not showing any radio variability. 
This result is consistent with the expectation that star formation occurs over extended temporal and spatial scales in galaxies, and therefore the integrated radio emission should not be variable on our timescales. This result is also a confirmation that our thresholds on $V$ and $\eta$ (described in \S \ref{sec:identifyvars}) are effective in separating variable sources from otherwise steady objects.


\subsubsection{Spectral Indices} 

\label{sec:alpha}
\begin{figure*}[!t]
\subfigure[\label{fig:alphaV}]{\includegraphics[width=0.5\textwidth]{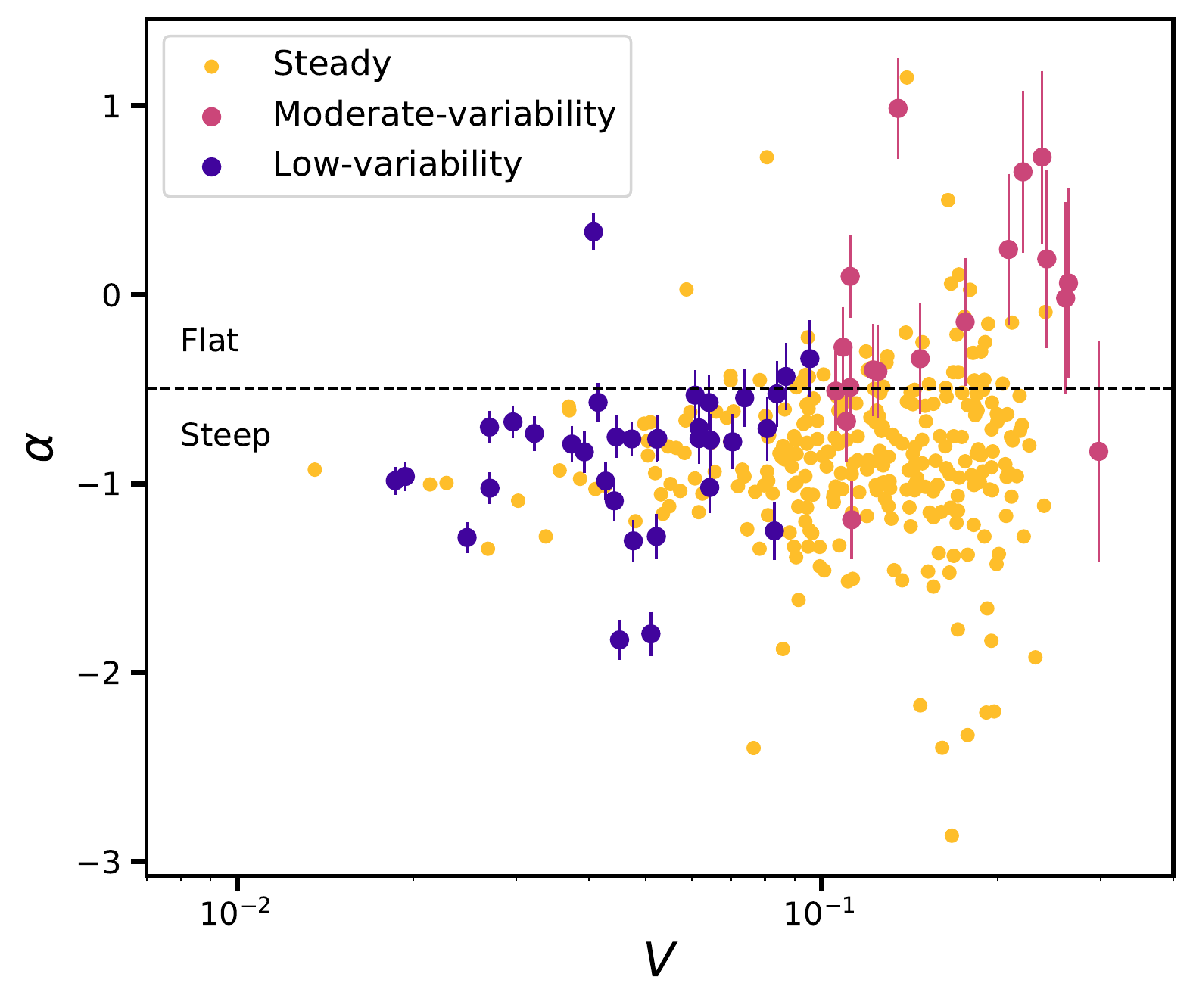}}
\subfigure[\label{fig:alphaeta}]{\includegraphics[width=0.5\textwidth]{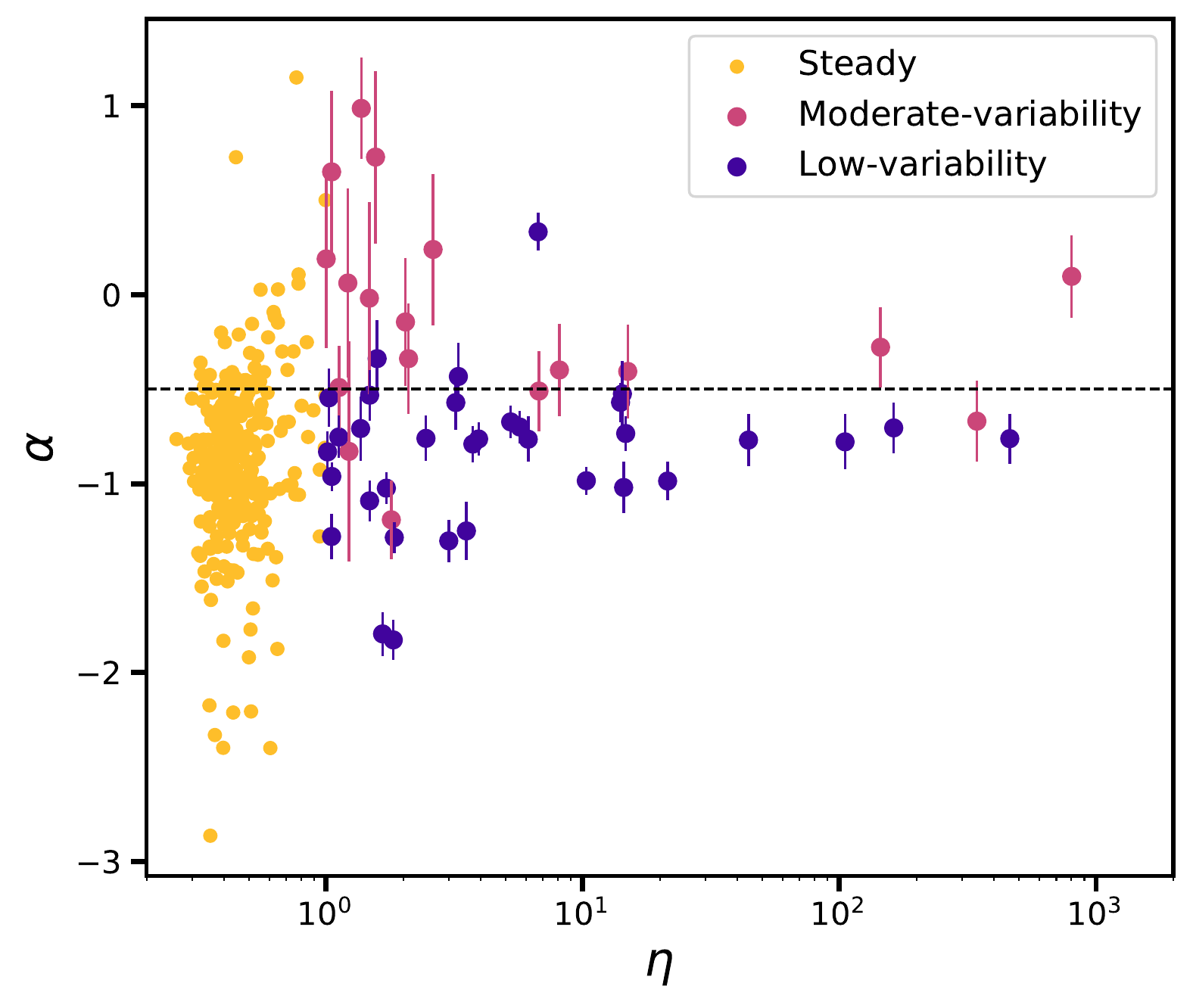}}
\caption{\textit{a)} 1.4--3 GHz spectral index $\alpha$ versus $V$ measured for our CHILES VERDES sources, color-coded by their variability category from Figure \ref{fig:Veta}.  \textit{b)} Same as in panel (a), except with $\alpha$ plotted against $\eta$ for our CHILES VERDES sources. We note that error bars for the steady sources are not shown for clarity, and error bars for the variable sources include the effects of variability.}
\label{fig:alpha}
\end{figure*}

For calculating spectral indices, we collect flux density information from the
VLA-COSMOS 1.4 GHz \citep{Schinnerer2010} and VLA-COSMOS 3 GHz surveys \citep{Smolcic2017a} surveys. Although these datasets have different image resolutions, the majority of our sources are point-like (as enforced in \S \ref{sec:chilesvarcandidates}).
The 1.4 GHz dataset consists of sources detected above 5$\sigma$ (RMS of 12 $\mu$Jy/beam) in the combined observations from the VLA-COSMOS Large and Deep Projects, while the 3 GHz dataset consists of sources detected above 5$\sigma$ (RMS of 2.3 $\mu$Jy). Both surveys provide the integrated flux densities (thus allowing for extended sources) and their uncertainties. 

While we could have been acquired spectral indices from the luminosities available in the S17 catalog, it did not have flux density uncertainties listed for the 1.4 GHz and 3 GHz measurements. We therefore cross-match our 370 CHILES VERDES sources with the 1.4 GHz and 3 GHz datasets, using the same 1$^{\prime\prime}$ radius circle criteria as in \S \ref{sec:topcat}.
We only select targets that were cross-matched to within 1$^{\prime\prime}$ for \emph{both} the 1.4 GHz and 3 GHz surveys, bringing us to a total of 339 sources. For a spectral index of the form $F_{\nu} \propto \nu^{\alpha}$, we calculate the uncertainty on $\alpha$ as:

\begin{equation}
\sigma_{\alpha} = \frac{\left[(\sigma_{F_1}/F_1)^2 + (\sigma_{F_2}/F_2)^2\right]^{1/2}}{\mathrm{ln}\left(\nu_1/\nu_2\right)}
\end{equation}
where $\nu_1$, $\nu_2$ are the observed frequencies (1.4 GHz and 3 GHz respectively), $F_1$, $F_2$ are the observed flux densities at these frequencies, and $\sigma_{F_1}, \sigma_{F_2}$ are the uncertainties on $F_1, F_2$. 
We assume that the uncertainties in the flux density measurements are due to statistical uncertainties ($\sigma_{F, st}$) as well as intrinsic variability of the source ($\sigma_{F, var}$), added in quadrature
\begin{equation}
\sigma_{F} = \sqrt{\sigma_{F, st}^2 + \sigma_{F, var}^2}
\end{equation}
The statistical uncertainties are listed in the catalogs. Since $V$ is the fractional RMS variation in the median flux density of our sources, we assume $\sigma_{F, var} = VF$. We are also assuming, for simplicity, that the variability $V$ at 3 GHz is the same as at 1.4 GHz for any source. 

The relation between $\alpha$ and variability $V$ of the 339 CHILES VERDES sources is shown in Figure \ref{fig:alpha}. We designate as ``flat spectrum" any source with $\alpha > -0.5$, consistent with the traditional definition \citep{Condon1984}. Steady sources show a range of spectral indices, with a mean and standard deviation of $\alpha = -0.88 \pm 0.45$, with nearly 97$\%$ of the sources being consistent with or below $\alpha=-0.5$ within their 1$\sigma$ uncertainties. This is consistent with the majority of steady sources being optically-thin synchrotron emitters. 

The variable sources, however, show a correlation between the spectral index and variability, with moderate-variability sources showing higher $\alpha$ on average compared to the low-variability sources (Figure \ref{fig:alphaV}). No such correlation is apparent between $\alpha$ and $\eta$, however (Figure \ref{fig:alphaeta}). We note that uncertainties in $\alpha$ are larger for sources with higher variability, because they are dominated by $V$. About 9 out of the 18 moderate-variability sources have a flat spectrum at the 1$\sigma$-level (i.e. median $\alpha$ is at least 1$\sigma$ above -0.5), and 4 have flat spectra at the 2$\sigma$ level. In comparison, only 1 out of the 34 low-variability sources (for which we obtained spectral index) has a flat spectral index (although it is flat at the 2$\sigma$ level). Such flat-spectrum sources could be core-dominated systems, systems with aligned jets such as blazars, and also young jets such as those associated with gigahertz-peaked spectrum sources \citep{ODea1998, Hovatta2014}.

\subsubsection{Redshifts}
\begin{figure}
    \centering
    \includegraphics[width=\columnwidth]{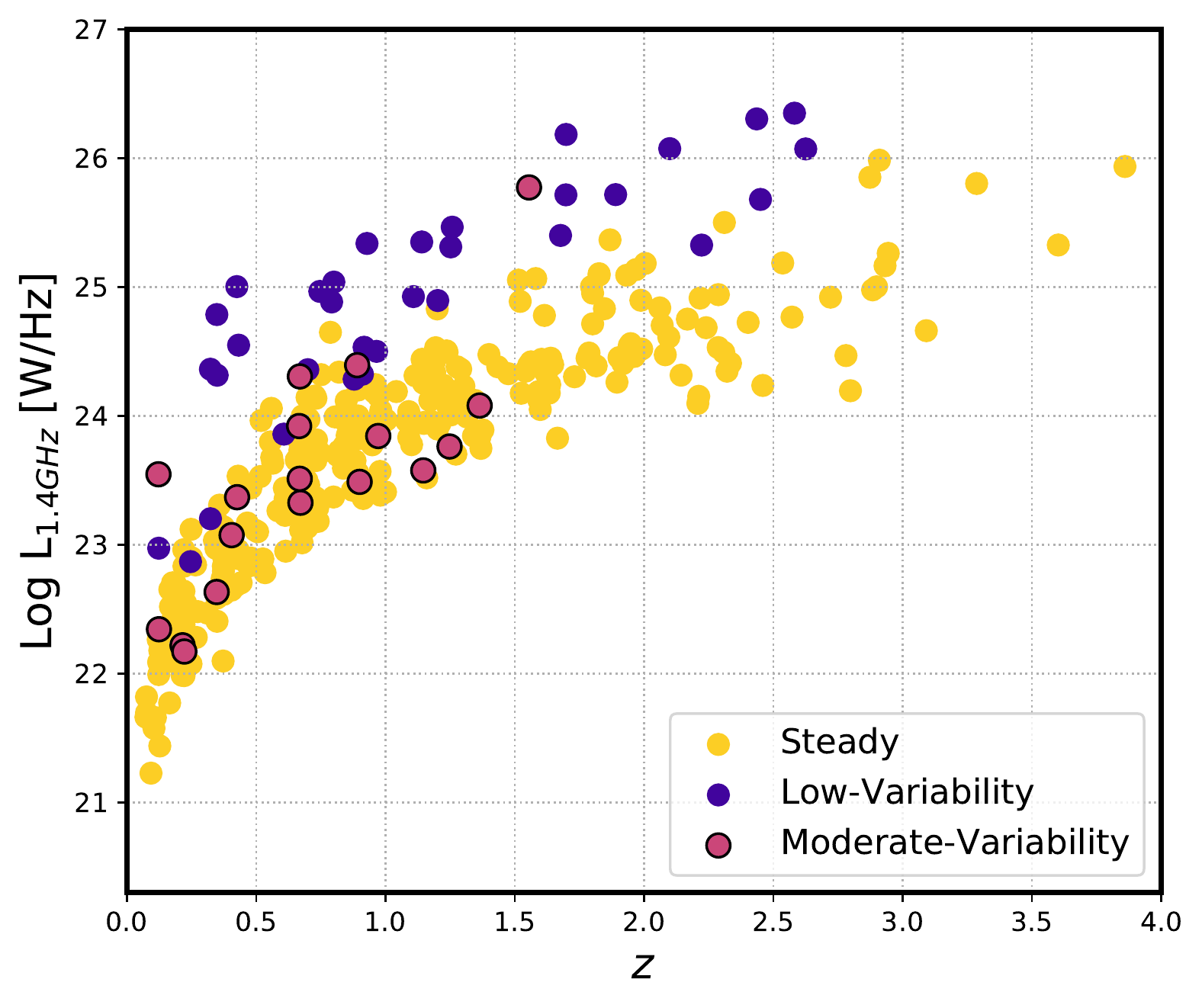}
    \caption{1.4 GHz luminosity versus redshift for CHILES VERDES sources, obtained from cross-matching with the S17 catalog. Source categories are labeled as usual.}
    \label{fig:threshold}
\end{figure}
One of the unique advantages of deep-drilling surveys is the opportunity to probe the cosmological evolution of sources. This is nicely demonstrated in the 
S17 catalog, which provides radio-selected sources with reliable redshifts out to $z\approx6$, sampling most of the cosmic history of star formation and black hole growth, and resolving the peak epochs at $z \approx 1.5-3$ \citep{Shankar2009, Madau2014, Heckman2014, Smolcic2017c}. 

Here we investigate the 1.4 GHz luminosity versus redshift distribution of CHILES VERDES sources. 
Redshifts for all sources were compiled by S17 from an exhaustive list of photometric and spectroscopic observations in the COSMOS field obtained both by the VLA-COSMOS 3 GHz team and archival studies, and we refer the reader to S17 and \cite{Delvecchio2017} for references and details of the measurements. The 1.4 GHz luminosities were measured by S17 using the rest-frame 1.4 GHz flux densities and spectral indices as discussed in \S \ref{sec:alpha}.


We show the 1.4 GHz luminosity-redshift distribution of our CHILES VERDES sources in Figure \ref{fig:threshold}. Our sources span nearly 6 orders of magnitude in luminosity (from $10^{21}-10^{27}$ W/Hz). Steady sources, being more numerous, span most of the observed range of redshifts between $z=0.07$ to $z=3.86$. moderate-variability sources are observed between $z=0.12-1.55$, while low-variability sources extend to higher redshifts, with maximum $z=2.63$. The luminosity evolution shows the characteristic shape of flux-limited surveys, whereby less luminous sources are discovered at lower redshifts. 
The distribution of radio luminosities at a given redshift range are similar to that of the steady sources. 
However, the moderate-variability sources are generally less luminous compared to the low-variability sources. The low-variability sources are among the most luminous sources in the sample across most of the observed redshift range. We note that compared to the full S17 sample, the CHILES VERDES sources 
are generally more luminous, mainly because of the shallower flux density limit of the individual CHILES VERDES epochs (3$\sigma$ limits in the range of 21-81 $\mu$Jy/beam) compared to 36 $\mu$Jy for VLA-COSMOS 1.4 GHz survey, and 6.9 $\mu$Jy for the VLA-COSMOS 3 GHz survey. Similarly, while the full S17 sample extends to $z\approx 6$, our catalog of steady sources is limited to $z<4$, since there are very few sources (only 95 out of the 8035 S17 sources with redshifts) beyond $z=4$.

\section{Discussion} \label{sec:discuss}
\subsection{Variability Statistics}
With 370 point-like sources within our CHILES VERDES search region, a tally of 18 moderate-variability sources suggest a variability fraction of $18/370 \approx 4.9\%$ (we exclude low-variability sources to  be consistent with previous radio-variability surveys that generally include sources with variability amplitude above 10$\%$). We estimate a Poisson statistical uncertainty $ = \sqrt{18}/370 \approx 1.1\%$. Our selection criteria may also add uncertainty to the variable fraction, but we expect this to be of the order of the Poisson uncertainty (e.g. if we relax our positional uncertainty criteria for point-like sources in \S \ref{sec:select} to $\Delta \theta_{pos} \leq 1.5^{\prime\prime}$, we get 616 sources and 21 moderate-variability sources, making the variability fraction $\approx 3.4\%$). For the mJy sources, we have 3 moderate-variability objects out of 32 point-like sources, giving a variability fraction for sources brighter than 1 mJy as $(9.3 \pm 5.4)\%$. In the sub-mJy regime, we have 15 moderate-variability objects out of 338 point-like sources, giving a variability fraction for sub-mJy sources $(4.4 \pm 1.1)\%$.  

These fractions are larger than most other 1.4 GHz surveys in the literature, where the typical variability fraction is constrained to less than 1$\%$ (e.g., \citealt{Croft2010, Thyagarajan2011, Mooley2013, Hancock2016, Bhandari2018}; similar statistics are echoed in other frequency bands as well).%
\footnote{We restrict our comparison to 1.4 GHz surveys for consistency, since variability can depend on the frequency band observed. The low variability fraction however has been reported in other frequencies as well, and a full list of studies is available here \url{http://www.tauceti.caltech.edu/kunal/radio-transient-surveys/}}. A caveat of comparison between previous variability surveys is the different sensitivities, timescales and variability thresholds used by studies. Additionally, because of our single pointing and effect of cosmic variance, the source counts may differ by about $10\%-35\%$ in another similar patch of the sky \citep{Heywood2013}. 

Regardless, our result agrees with previous findings in that only a small fraction of the extragalactic radio population exhibits any statistically significant variability. Our survey is likely picking up a variability fraction $> 1\%$ because of the combination of depth (which helps probe the rich extragalactic population between 0.1--1 mJy at $>5\sigma$), cadence (allowing us to resolve weekly--monthly timescale variability), and survey length (making our survey also sensitive to slow year-long variability that appears to be common in AGN). Previous studies have hinted at the possible dependence of variability fraction on factors like sensitivity \citep{Carilli2003, Thyagarajan2011, Hancock2016, Radcliffe2019}, cadence and timescales probed \citep{Hodge2013, Radcliffe2019}.

We also found that no variable source brightened by more than a factor of 2 during our course of observations, which is consistent with the previously mentioned deep-field studies \citep[e.g.,][]{Mooley2013, Hancock2016, Bhandari2018, Radcliffe2019}, implying that dramatic brightening events, such as those expected from flares and renewed jet activity in AGN discovered in wider-area surveys \citep{Mooley2016}, are relatively rare. Given the 370 variability candidates recovered by our pipeline, and an observing baseline of $\sim 5.5$ years, we estimate that an object brightening by more than a factor of two happens at a rate of less than twice per square degree per 1000 years. 

\subsection{Sources of variability observed in CHILES VERDES} \label{sec:sourceofvars}
From Figures \ref{fig:rlcsfvars1}, \ref{fig:rlcsfvars2} and \ref{fig:rlcsflowvars}, it is clear that radio variables in CHILES VERDES vary on a range of timescales, from a few days/week to years, and that individual sources themselves may be associated with variability on multiple timescales (e.g., CV17). 

Variability in radio sources has been attributed to both extrinsic effects (such as scintillation by the Galactic ISM) and intrinsic effects (i.e., arising from shocks in the black hole jet). We can place a rough estimate on the contribution from scintillation using the models of \cite{Walker1998} and the distribution of ISM electron density from \cite{Cordes2002}. \cite{Walker1998} parameterizes the amount of scattering induced in a radio wavefront by the turbulent ISM (approximated as a thin phase-scattering screen) in terms of the scattering strength $\zeta = (\nu_0/\nu)^{17/10}$, where $\nu$ is the observing frequency and $\nu_0$ is the line-of-sight-dependent transition frequency between weak ($\zeta\ll1$) and strong ($\zeta\gg1$) scattering regimes. Physically, broad-band scintillations are greatest when $\nu=\nu_0$, and therefore $\zeta=1$, implying a modulation index (corresponding to $V$) = 1. In the direction of the CHILES field, $\nu_0 = 9.4$ GHz\footnote{based on \url{https://www.nrl.navy.mil/rsd/RORF/ne2001/}}, which means $\zeta \approx 24 \gg1$ for $\nu = 1.45$ GHz (our observing frequency), putting us in the strong scattering regime. In this regime, scintillation can be refractive or diffractive, but only the former will be captured in broad-band observations. For refractive scintillation, the modulation index $V = (\nu/\nu_0)^{17/30} \approx 0.35$ and the timescale of variation $t_r \sim (2\mathrm{hr})(\nu_0/\nu)^{11/5} \approx 5$ days for $\nu = 1.45$ GHz (marked as an orange dashed line in the SFs plotted in Figures \ref{fig:rlcsfvars1}, \ref{fig:rlcsfvars2} and \ref{fig:rlcsflowvars}). Hints of a slope break on timescales $\sim t_r$ are seen in the SFs of some of the variables, particularly prominent in the bright sources CV16 and CV17 (Figure \ref{fig:rlcsfvars2}), the fainter source CV4 (Figure \ref{fig:rlcsfvars1}), as well as the low-variability sources CV38 and CV50 (Figure \ref{fig:rlcsflowvars}). Several sources show variability on timescales (based on plateaus in their SFs) almost an order of magnitude above the scintillation timescale, such as CV3, CV5, CV6, CV10, CV13, CV17 and CV18, along with some of the low-variability sources. Such variability on longer timescales are therefore not likely entirely driven by scintillation.

Another constraint on the variability timescale comes from the fact that the brightness temperature for incoherent synchrotron sources is limited to $<3\times10^{11}$ K due to inverse Compton losses \citep{Kellermann1969, Readhead1994}. The brightness temperature depends on the source angular extent ($\theta$) as $T_B \propto \theta^{-2}$. By a light travel-time argument, $\theta \leq c \tau$, where $\tau$ is the variability timescale of the source, putting a lower-limit on $T_B$ \citep{Pietka2015}. However, this lower limit needs to be below the inverse Compton limit, which puts a physical lower limit on $\tau$. Following \cite{Bhandari2018} Eq.\ (4), this limit is defined as:
\begin{equation}
\tau_{min} \geq \sqrt{\frac{\Delta F D^2}{2 k_B \nu^2 T_{B,max}}}
\end{equation}
where $k_B$ is the Boltzmann constant, $\Delta F$ is the amplitude of flux density variation, $D$ is the luminosity distance to the source, $\nu$ is the observing frequency, and $T_{B,max}=3\times 10^{11}$ K is the maximum allowed brightness temperature. Note that this does not include any correction for Doppler boosting effects. Assuming $\Delta F = V \bar{I}$, we show $\tau_{min}$ in the SFs of our variables (green dashed lines on the SFs in Figures \ref{fig:rlcsfvars1}, \ref{fig:rlcsfvars2} and \ref{fig:rlcsflowvars}). We find $\tau_{min}$ larger than characteristic timescales in many of the sources like CV6, CV10, CV11, CV13, and CV18, and in all the low-variability sources in Figure \ref{fig:rlcsflowvars}. If a source is varying on these timescales, it would exceed the brightness temperature limit. 

This implies that the aforementioned monthly--yearly variability are either driven by long-timescale scintillation, or by Doppler-boosted intrinsic variability. The scintillation timescale derived earlier assumes that the extent of radio emission from sources ($\theta_s$) is smaller than the first Fresnel zone $\theta_r$ (the characteristic scale over which 1 rad of phase-delay is induced in the wavefront). However, if components larger than the central engine (e.g. star-forming regions, lobes) also have significant contribution to the radio emission, then $\theta_s$ could be larger than $\theta_r$, and scintillation timescale becomes larger by a factor ($\theta_s/\theta_r$). However the modulation index on these timescales will also be decreased by a factor $(\theta_r/\theta_s)^{7/6}$, whereas from the SFs (e.g. CV18 or CV13) we see that longer timescales have more power than shorter ones. For scintillation to operate on these timescales in our 1.4 GHz observations would either need significant departures from the \cite{Walker1998} model assumptions or the \cite{Cordes2002} electron distribution data along our line-of-sight. The other possibility is that the variability is intrinsic, but affected by Doppler boosting, which increases $T_B$ for a given source extent by a multiplicative factor $\delta^3$ that depends on the jet kinematics \citep{Cohen2003, Kellermann2007, Hovatta2009}. Doppler boosting has been observed in sources even for modest inclinations \citep{Zensus1997}. Further work obtaining VLBI measurements of the CHILES field \citep[e.g.,][]{HerreraRuiz2017} can presumably help separate the contributions of scintillation and intrinsic variation.

\subsection{Reflection on Future Surveys} \label{sec:reflections}

\begin{figure}
    \centering
    \includegraphics[width=\columnwidth]{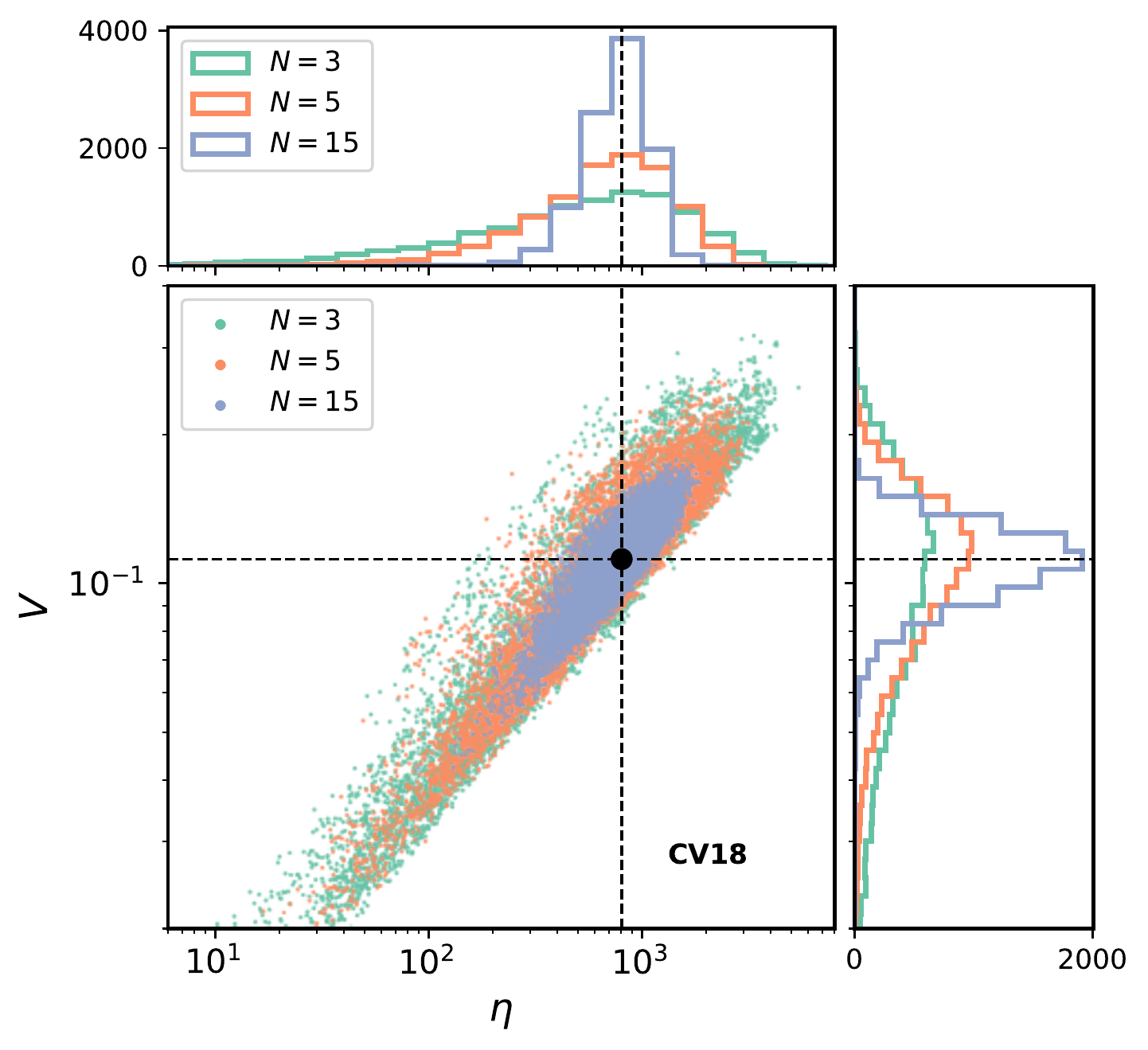}
    \caption{Effect of sparse sampling demonstrated by the $V-\eta$ distribution for 10$^4$ randomly drawn flux densities from the light curve of CV18 in sets of 3, 5 and 15 epochs (shown in colors). The top and side histogram shows the distribution of $V$ and $\eta$ respectively for the randomly drawn points (See \S \ref{sec:reflections} for details). Black circle with cross-hairs show the measured $V$ and $\eta$ for CV18 from all 172 epochs.}
    \label{fig:samplingVeta}
\end{figure}

Our work underscores the importance of deep-field surveys in the future for studying transients and variables. While wide-field surveys 
discover more transient events, the ability of deep-field surveys to sample radio variability with more epochs for the same time-allocation is crucial for characterizing the temporal behavior of the radio sky. This is particularly important for characterizing extragalactic variables that show flux density variations on a range of timescales; sampling a few epochs can lead to less precise determination of their variability. As an example, we randomly sample 10$^4$ sets of $N=3, 5, 15$ epochs from the light curve of CV18 (mimicking the epoch intervals of previous studies), and calculate $V$ and $\eta$ for each set (shown in Figure \ref{fig:samplingVeta}). In all cases, the mean $V$ and $\eta$ are similar to the measured values from 172 epochs, but the percentage uncertainty (standard deviation divided by mean) in the measured $V$ and $\eta$ is 51$\%$ and $93\%$ respectively for 3 epochs, and drops to 18$\%$ and $33\%$ respectively for 15 epochs. Statistically therefore there is a higher chance of over/underestimating the significance and variability of a source with only a few epochs. Although the number of epochs per source depends on the survey area and total time allocation, it would be advisable to have $\gtrsim$10 epochs to accurately characterize the parameter space of variables, and by extension, steady sources as well as bright transients.

\section{Conclusion}\label{sec:conclusion}
We have conducted a unique deep-field radio variability survey in a 0.44 deg$^2$ area of the data-rich COSMOS field using 1--2 GHz VLA continuum observations. Our sources were observed from October 2013 to April 2019, sampled with 172 epochs with RMS noise in the range of $7-27\, \mu$Jy and synthesized beam resolution of $\sim 4.3^{\prime\prime}$. On average, the cadence was $<2$ day (except for the year-long gaps between each B-configuration semester) and the observing time totals almost 960 hours of on-source time. 

A total of 370 point-like sources were detected using TraP,
of which 18 were designated `moderate-variability' ($>$10\% variability), 40 were designated `low-variability sources' ($<$10\%, but still significant, variability), and the rest as `steady' sources (no significant variability). 
Classification of these sources was carried out using the coefficient of variation $V$ (equivalent to RMS fractional flux density variation, or the modulation index) and $\eta$ (the reduced $\chi^2$ statistic). 
The fraction of 
$>$80 $\mu$Jy radio sources that are variable at the level of 10$\%$ or more is 4.9$\%$.
We characterized our sources using their radio light curves, SFs, spectral indices obtained by matching VLA-COSMOS 1.4 GHz and 3 GHz catalogs, and multi-wavelength characteristics obtained by cross-matching with \cite{Smolcic2017b} catalog. This catalog provided redshifts, intrinsic luminosities, and host-galaxy classifications for VLA-COSMOS 3 GHz sources by cross-matching with archival photometric, spectroscopic and multi-wavelength catalogs. Radio sources were classified as 
HLAGN, MLAGN, and star-forming galaxies based on thresholds in X-ray luminosity, mid-IR colors, optical-to-mm SEDs, 
and radio excess over expectations from IR-derived star-formation rates. 

The 18 moderate-variability sources show complex time-series behavior, including slow month--year long variability (e.g., CV17, CV18), to faster day--week timescale variability (e.g., CV16). Flux density variations are observed at the level of $\gtrsim 10\%$ for the brightest (mJy) sources and up to $\sim 30\%$ in the sub-mJy regime. SFs of these sources show the expected plateaus at short timescales due to measurement errors, and at long timescales in the white-noise/flicker noise regime of variability (roughly equal to the variance of flux densities). The intermediate timescales show a variety of shapes and breaks, possibly indicating the superposition of processes occurring at multiple characteristic timescales. We note that such detailed SFs were possible in a blind survey due to the large number of epochs sampling the light curves. 

The moderate-variability sources generally have a maximum-to-median flux density ratio between 1.3--2, indicating brighter flaring events in AGN are quite rare. Host galaxies for all the moderate-variability sources have AGN signatures, the majority of which are also associated with radio excess, which further corroborates the presence of a relativistic jet producing radio emission. The host galaxies are a mixture of HLAGN and MLAGN
spanning a redshift range of z=0.22--1.56.
About half of the moderate-variability sources are consistent with having flat spectra in the 1--4 GHz range ($F_{\nu} \propto \nu^{\alpha}$, with $\alpha>-0.5$), and are candidates for core-dominated systems, blazars and/or young jets associated with GHz-peaked spectra. 

Most of the 40 low-variability sources show flux density variation at levels of $2-8\%$. Physically there are similarities between the host galaxy properties of low- and moderate-variability sources, in that they are associated with HLAGN and MLAGN in similar proportions. 
However, low-variability sources are generally brighter and more luminous at radio wavelengths, extend to farther distances ($z \approx 2.8$), have smaller maximum-to-median flux density ratios ($\lesssim 1.5$), and have generally steeper 1--4 GHz spectra.  A few interesting objects (e.g., CV55, CV50 and CV57) show variability patterns similar to our objects classified as moderate-variability sources, indicating that $V-\eta$ thresholds can miss interesting objects, and it is important to manually inspect the light curves of sources near such thresholds.

Steady sources form the bulk of our time-series sample. The largest group is star-forming galaxies (45$\%$), 
and the rest of the sample contains roughly equal proportions of HLAGN and MLAGN. This indicates that not all AGN are associated with radio variability (at least with the cadence, depths and duration explored in this paper). 

For the 58 CHILES VERDES sources that demonstrate significant variability, their light curves can be explained as a combination of scintillation at short timescales and intrinsic Doppler-boosted variability on longer timescales. Our observing frequency, bandwidth and line-of-sight off the Galactic plane implies any scintillation would be refractive, with variability at the level of $<35\%$ with a characteristic timescale $\sim 5$ days based on the \cite{Walker1998} model. Several variables show breaks in their SFs at these timescales indicating the role of scintillation. However, breaks at longer timescales of months and years are also observed, indicating intrinsic AGN variability. However, these observed timescales are lower than the minimum timescale needed to remain below the inverse Compton brightness temperature limit $3\times10^{11}$ K. This implies that scintillation still dominates at longer timescales in excess of model predictions, or that the variability is intrinsic but Doppler-boosted.

\section*{Acknowledgements}

We are grateful to Jacqueline van Gorkom for her leadership of the CHILES survey, Ian Heywood for discussions about source counts, Vernesa Smolcic and Ivan Delvecchio for guidance with the VLA COSMOS 3 GHz catalog, and Dillon Dong and Kristina Nyland for discussions about radio variability. The National Radio Astronomy Observatory is a facility of the National Science Foundation operated under cooperative agreement by Associated Universities, Inc. This paper includes observations obtained at the Southern Astrophysical Research (SOAR) telescope, which is a joint project of the Ministério da Ciência, Tecnologia, Inovações e Comunicações do Brasil (MCTIC/LNA), the US National Science Foundation’s NSF’s NOIRLab (NOIRLab), the University of North Carolina at Chapel Hill (UNC), and Michigan State University (MSU). This research has made use of the NASA/IPAC Infrared Science Archive, which is funded by the National Aeronautics and Space Administration and operated by the California Institute of Technology. This research has made use of NASA’s Astrophysics Data System Bibliographic Services.

S.S., E.T., and L.C.\ are grateful for support from NSF grants AST-1412549 and AST-1907790. J.S.\ and S.S.\ acknowledge support from the Packard Foundation. C.P. and E.W. are grateful for support from the NSF grants AST-1413099.

\vspace{5mm}
\facilities{VLA, SOAR, HST, IRSA}


\software{numpy \citep{numpy}, scipy \citep{scipy}, matplotlib \citep{matplotlib}, astropy \citep{astropy1, astropy2}, CASA \citep{CASA}}

\bibliographystyle{aasjournal}
\bibliography{CV_Variables_Final.bib}
\appendix
\section{Redshift determination for CV17}
\begin{figure}
    \centering
    \includegraphics[width=0.6\textwidth]{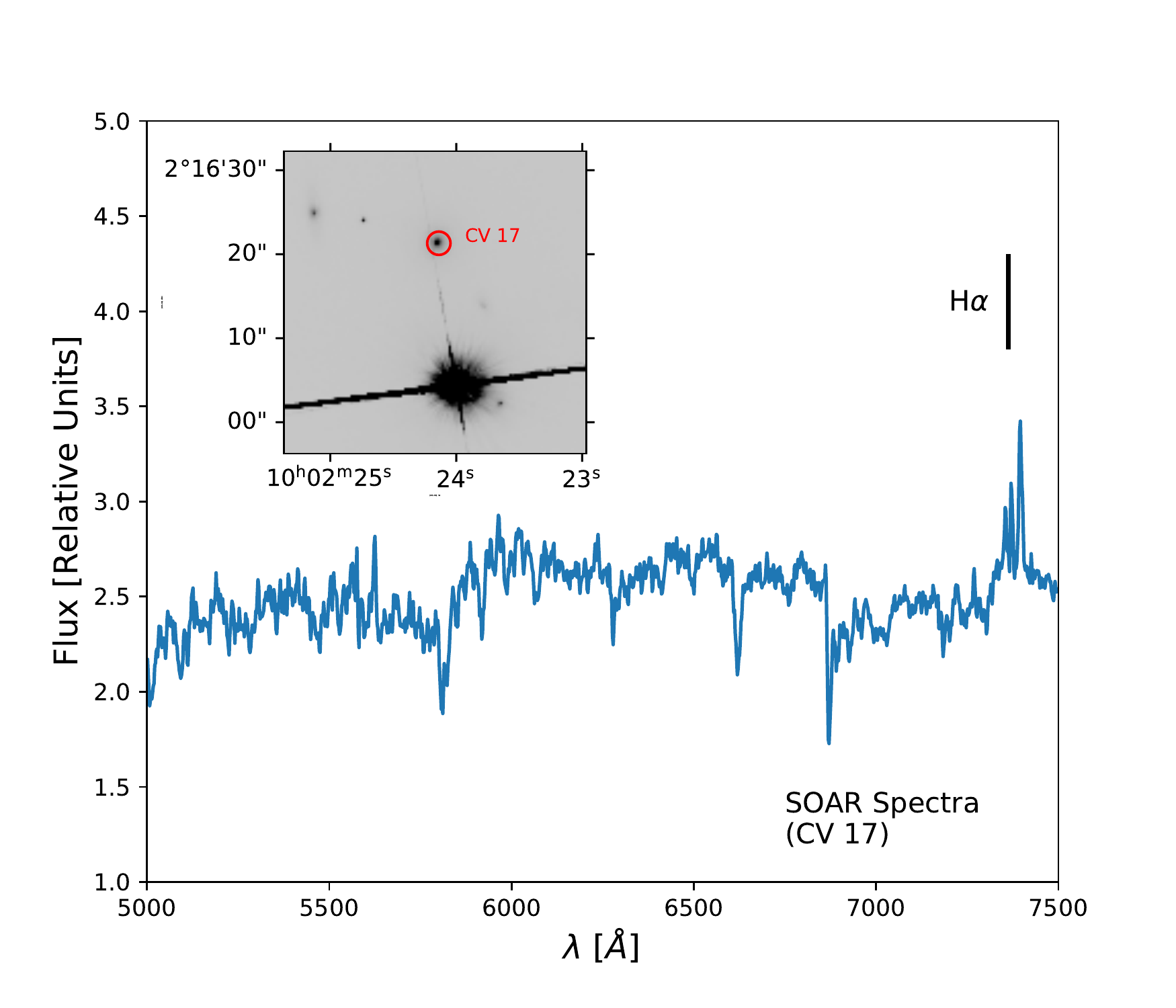}
    \caption{SOAR spectrum of CV17 showing the location of redshifted H$\alpha$ + [\ion{N}{2}]. The inset shows an \emph{HST}/ACS image of CV17 and the bright foreground object.}
    \label{fig:43031cutout}
\end{figure}
CV17 is one of the brightest variable objects, but did not have redshift listed in the COSMOS catalogs. This is likely due to its location in a masked area in optical data because of a bright foreground source (Figure \ref{fig:43031cutout}). We obtained an optical spectrum of CV17 with the Goodman Spectrograph \citep{Clemens04} on the SOAR telescope on 2019 Nov 4. Two 20-min exposures were obtained, using the 400 l mm$^{-1}$ grating and covering a wavelength range of $\sim 3850$--7850 \AA\ at a resolution of about 5.9 \AA. The spectra were optimally extracted in the usual way and combined, with a first-order flux calibration applied. The spectrum is shown in Figure \ref{fig:43031cutout}. It shows characteristic absorption lines of an intermediate-to-old stellar population as well as narrow emission lines. By cross-correlating the emission lines with an emission-line galaxy template from XCSAO\footnote{http://tdc-www.harvard.edu/iraf/rvsao/Templates}, we find $z=0.122$.

\end{document}